\documentclass[pre,twocolumn,showpacs,amsfonts,amssymb]{revtex4}
\usepackage{amsmath}
\usepackage{graphics}
\usepackage[pdftex]{graphicx}
\usepackage{multirow}

\newcommand{\I}{{\rm I}}
\newcommand{\II}{{\rm II}}

\begin{document}

\title{Monte Carlo Approaches for Simulating a Particle at a Diffusivity Interface and the ``Ito--Stratonovich Dilemma''}
\date{\today}
\author{Hendrick W. \surname{de Haan}}
\affiliation{Department of Physics, University of Ottawa, 150 Louis-Pasteur, Ottawa, Ontario, Canada, K1N 6N5}
\author{Mykyta V. Chubynsky}
\affiliation{Department of Physics, University of Ottawa, 150 Louis-Pasteur, Ottawa, Ontario, Canada, K1N 6N5}
\author{Gary W. Slater}
\affiliation{Department of Physics, University of Ottawa, 150 Louis-Pasteur, Ottawa, Ontario, Canada, K1N 6N5}
\email{gary.slater@uottawa.ca}

\begin{abstract}
The possibility of different interpretations of the stochastic term (or calculi) in the overdamped Langevin equation for the motion of a particle in an inhomogeneous medium is often referred to as the ``Ito--Stratonovich dilemma,'' although there is, in fact, a continuum of choices. We introduce two Monte Carlo (MC) simulation approaches for studying such systems, with both approaches giving the choice between different interpretations (in particular, Ito, Stratonovich, and ``isothermal''). To demonstrate these approaches, we study the diffusion on a 1D interval of a particle released at an interface (in the middle of the system) between two media where this particle has different diffusivities (for example, two fluids with different viscosities). We consider the properties of the particle distribution for reflecting boundary conditions at the ends of the 1D interval. A discontinuity at the interface in the stationary-state particle distribution is found, except for the isothermal case, as expected. We also study the first-passage problem using absorbing boundary conditions. Good agreement is found when comparing the MC approaches against theoretical predictions as well as Brownian and Langevin dynamics simulations. Additionally, while this problem was chosen primarily to verify the algorithms, the results themselves turn out to be interesting --- particularly when comparing across interpretations. For instance, we report that: 1) for some calculi, there can be more particles on the low-viscosity side at earlier times and then more particles on the high-viscosity side at later times; 2) there is no preference to end up on a particular wall for the Ito variant, but a bias towards the wall on the low-viscosity side in all other cases; 3) the mean first-passage time to the wall on the low-viscosity side grows as the viscosity on the high-viscosity side is increased, except for the isothermal case where it approaches a constant; 4) when the viscosity ratio is high, the first-passage-time distribution for the wall on the lower-viscosity side is much broader than for the other wall, with a power-law dependence of the former in a certain time interval whose exponent depends on the calculus; 5) the average portion of time the particle spends on a particular side can be very different from the probability to reach the wall on that side and depends significantly on how the averaging is done.

\end{abstract}

\pacs{02.70.Tt, 05.40.Jc, 05.10.Ln, 05.10.Gg}

\maketitle

\section{Introduction}
\label{sec:intro}
While deceptively simple to consider conceptually, 
the diffusion of a particle in an inhomogeneous medium is a subtle problem \cite{lancon2002,vanmilligen2005, lau2007, bringuier2011}.
For a system consisting of a high viscosity and a low viscosity region, will a tracer particle be found more often on the ``thicker" side, the ``thinner" side, or with an equal probability at all locations?
While an assumption of a Boltzmann distribution would suggest an equal probability at all locations, 
a simple physical argument suggests that the particle may be more likely to be found on the high viscosity side.
If the only inhomogeneity in the system is the viscosity, then statistically, the dynamics are the same at all points with the exception that time is rescaled on the high viscosity side where the particle moves slower.
Hence the particle will spend a greater amount of time on that side and should be more likely to be found there.

While this discussion can go back and forth, in truth each answer corresponds to a different formulation of the problem --- both with their own mathematical and physical assumptions. In essence, the question boils down to considering a particle at the viscosity interface: will there be an additional ``push'' at the interface to the low viscosity side that will compensate the ``trapping'' in the high viscosity region due to the slower dynamics there? Let us assume that the fluids can be considered implicitly, so their effect on the diffusing particle is reduced to a random force and a friction force that the particle feels.
We further assume that the interface is sharp and that there are no interfacial effects such as surface tension that would make the particle prefer or avoid the interface. 
Finally, we assume the particle motion is overdamped such that it can be described by the \textit{overdamped Langevin equation} \cite{coffey2005}. 
The situation where the assumption of equivalent dynamics (apart from the time scale) holds and particles accumulate on the high viscosity side corresponds to interpreting the stochastic term in the overdamped Langevin equation according to the rules of the \textit{Ito calculus} \cite{ito1951}.
In other interpretations, of which the most well-known corresponds to \textit{Stratonovich calculus} \cite{stratonovich1966}, but perhaps of most interest to physicists is the so-called \textit{isothermal} formulation \cite{hanggi1978,hanggi1980,hanggi1982,klimontovich1990}, the particle ``feels" the interface and has a tendency to move to the low viscosity side.
In particular, in the isothermal case (also referred to in the literature as H\"{a}nggi-Klimontovich or kinetic \cite{sokolov2010}) this bias at the interface cancels the trapping effect and, in the stationary state, the particle has an equal probability of being found at all locations.

There have been several experimental explorations of systems with nonuniform viscosity or, more generally, particle diffusivity \cite{vanmilligen2005, lancon2002}.
Most relevantly, van Milligen \textit{et al.} \cite{vanmilligen2005} performed experiments in which 
the relative concentration of dye was monitored in a system comprised of two viscosity regions created by adding gelatine to one half of it.
In this setup, the dye was found to accumulate on the high viscosity side.
As demonstrated in their paper, the concentration distributions they observed can be approximated by solutions of the Fokker-Planck equation, which corresponds to the Ito formulation (see Sec.~\ref{sec:theory}). 
On the other hand, in experiments by Lan\c{c}on \textit{et al.} \cite{lancon2002}, particles diffused between two nearly, but not perfectly, parallel walls, which, as the authors showed, is equivalent to diffusion with a variable diffusion coefficient. In contrast to Ref.~\cite{vanmilligen2005},
while individual particles were observed to drift to regions of higher diffusivity, there was no overall flux when starting with a uniform concentration of particles and hence the uniform concentration was preserved, which is consistent with the isothermal case.
On the theoretical side, the general conclusion has been that different formulations can be appropriate depending on the details of the system. For instance, in a simple model of diffusion in a slowly modulated periodic potential by Sokolov~\cite{sokolov2010}, Ito, Stratonovich, and isothermal rules correspond to different variants of the model.

Given the experimental and theoretical relevance of diffusion in inhomogeneous media, it is desirable to be able to model it computationally. One popular computational approach to modeling the dynamics of particles in a fluid medium is molecular dynamics (MD) \cite{frenkel}. 
Two of the most common MD approaches assuming an implicit fluid are Brownian dynamics (BD) and Langevin dynamics (LD). The BD approach consists in solving numerically the overdamped Langevin equation in the Ito formulation. LD drops the overdamped assumption and adds an inertial term to the equation of motion that is then solved numerically. Interestingly, even in the limit of an infinitely large damping (but decreasing the time step so it is always much smaller than the corresponding relaxation time), LD is not reduced to BD and instead corresponds to the isothermal formulation of the overdamped Langevin equation. Thus the choice between the two methods is not arbitrary when an inhomogeneous system is considered.

\textit{Monte Carlo (MC) methods} are another frequently used group of computational approaches \cite{barkema}. Unlike MD, MC methods generally sacrifice at least some aspects of the dynamics of the system to achieve a computational speedup compared to MD. How much is sacrificed can vary. The popular Metropolis algorithm \cite{metropolis1953} is designed solely to reproduce the equilibrium state and is generally unsuitable for studying the dynamics, even qualitatively. For instance, in the problem of inhomogeneous diffusion considered here, a particle would diffuse equally rapidly regardless of the viscosity, \textit{unless} care is taken to adjust the step length and/or the time increment at every MC step (the \textit{time step}) appropriately, which is not a part of the basic algorithm. However, there are also methods (\textit{dynamical MC}) which do reproduce the principal aspects of the dynamics and even the correct time scale of the process. This includes generally faster, but sometimes less accurate and/or reliable methods where the particles are restricted to pre-defined points in space forming a lattice (\textit{lattice MC}, or LMC), as well as off-lattice methods. Dynamical LMC methods include kinetic Monte Carlo 
\cite{bortz1975,fichthorn1991,battaile2002}
and methods designed to reproduce the correct dynamics in an arbitrarily strong uniform field \cite{gauthier2004,gauthier2005,gauthier2008}.

In this article, we present two dynamical LMC algorithms which can be used to generate results consistent with different formulations of the inhomogeneous diffusion problem. We consider a simple system consisting of a single random walker diffusing in a box which contains two separated fluids in equal proportions:
a fluid of viscosity $\eta_L$ on the left side and viscosity $\eta_R$ on the right side, similar to the experimental setup in Ref.~\cite{vanmilligen2005}.
Hence, there is a viscosity interface at the center of the box. Projecting the motion of the random walker onto the direction perpendicular to the interface reduces the problem to a 1D one. The particular choice of the problem with a sharp interface has been made for two (in a way, opposite) reasons. On the one hand, \textit{a priori} it is the simplest setup in which to study diffusion in an inhomogeneous medium; it lets us define simple quantities characterizing the process, such as the fraction of the time spent on a particular side of the interface. On the other hand, when it comes to actually solving the problem, it turns out there are difficulties specifically in the case of a sharp interface, as discussed in Sec.~\ref{sec:theory}, so in that sense this problem is, in fact, \textit{more} complicated than that of a gradual change in viscosity; our algorithms should generalize straightforwardly to the latter case. In addition, in future work we intend to apply the algorithms developed here to studying translocation of a polymer through a nanopore in a thin membrane, in which case having different viscosities on the two sides of the membrane with a sharp boundary between them corresponding to the pore is natural.
Overall, the system of a single random walker in a fluid with a single, sharp viscosity interface is chosen as a simple ``toy problem" such that we can clearly demonstrate the validity of our LMC algorithms.
However, as will be shown in the results section, many interesting features arise for the dynamics of a particle in such a system --- particularly when contrasting results for different calculi.
This manuscript thus not only introduces LMC approaches for simulating diffusion in a system containing regions of differing diffusivity, 
but it demonstrates the complexity which arises for even the simplest case and highlights the importance of choosing the calculus that is appropriate for the problem under study.

In the next section, we briefly review the relevant parts of the theory of the overdamped Langevin equation for the motion of a Brownian particle and obtain the corresponding equation for the particle concentration. We then introduce Ito, Stratonovich, and isothermal calculi giving some examples of their physical relevance. We also point out some problems with using the overdamped Langevin equation in the case of a sharp interface. In the following two sections, we consider different simulation techniques that can be used for the problem of particle diffusion. We first note that the well-known BD and LD methods correspond to Ito and isothermal calculi, respectively, and then introduce two MC algorithms. The details of the simulation protocol are outlined in Sec.~V, after which in Sec.~VI we present comparisons between different methods (BD, LD, MC) and results of the simulations. We end the paper with a discussion of possible uses of the methods developed here and their future generalizations.

\section{Theory}
\label{sec:theory}

\subsection{Diffusion in a liquid}
Diffusion of a pointlike particle along the $x$ axis under the influence of a potential $U(x)$ can be described by the Langevin equation \cite{pottier2009},
\begin{equation}
m\frac{d^2}{dt^2}x=-\zeta \frac{d}{dt} x - \frac{d}{dx} U(x) + \sqrt{2 kT \zeta} R(t),\label{LDiner}
\end{equation}
where $m$ is the particle mass, $k$ is the Boltzmann constant, $T$ is the temperature, and $R(t)$ is a random function that satisfies $\langle R(t) \rangle =0$ and $\langle R(t) R(t') \rangle = \delta(t-t')$, where $\langle\ldots\rangle$ denotes the ensemble average. The coefficient of the last term follows from the fluctuation-dissipation theorem. It is assumed that the drag force $F_f$ (the first term on the right-hand side) is proportional to the particle velocity $v$,
\begin{equation}
F_f=-\zeta v,
\label{ffric}
\end{equation}
where $\zeta$ is the friction coefficient. For a particle moving in a fluid, $\zeta$ is proportional to the fluid viscosity $\eta$:
\begin{equation}
\zeta=\kappa\eta,\label{zetavseta}
\end{equation}
where $\kappa$ is constant for a given particle. For instance, for a spherical particle of radius ${\cal R}$, according to the Stokes' law (which assumes no-slip boundary conditions for the fluid),
\begin{equation}
\kappa=6\pi {\cal R};
\end{equation}
for other particle shapes and boundary conditions the expression for $\kappa$ will be different, but Eq.~(\ref{zetavseta}) is still valid.

In many cases of interest, the left-hand side (or the inertial term) in Eq.~(\ref{LDiner}) is small. If this term is neglected (the \textit{overdamped limit}), we obtain the \textit{overdamped Langevin equation}
\begin{equation}
\zeta \frac{d}{dt} x = - \frac{d}{dx} U(x) + \sqrt{2 kT \zeta} R(t).
\label{LDBM}
\end{equation}
It is this limit that we will consider in what follows. However, neglecting the inertial term can bring about subtle problems, as discussed below.

Taking the case of free diffusion [$U(x)=0$] we can rewrite Eq.~(\ref{LDBM}) as
\begin{equation}
\frac{d}{dt} x = \sqrt{\frac{2 kT}{\zeta}} R(t).\label{SDE}
\end{equation}
The particle diffusion coefficient (or diffusivity) $D$ is defined by the expression
\begin{equation}
\langle \Delta x^2 \rangle=2D\Delta t,\label{D}
\end{equation}
where $\Delta x$ is the particle displacement over time $\Delta t$. Based on Eq.~(\ref{SDE}), it can be shown that
\begin{equation}
D=\frac{kT}{\zeta}=\frac{kT}{\kappa\eta}.
\label{diffzeta}
\end{equation}
We then obtain
\begin{equation}
\frac{d}{dt} x = \sqrt{2 D} R(t).\label{SDE_Dconst}
\end{equation}

\subsection{Inhomogeneous media}
The consideration below and in Section~\ref{fluxes} is similar to Ref.~\cite{lancon2002}, but somewhat more detailed.

If we consider diffusion in a stationary inhomogeneous medium, $D$ will be dependent on $x$ such that
\begin{equation}
\frac{d}{dt} x = \sqrt{2 D(x)} R(t).\label{SDE_D}
\end{equation}

To proceed, we must determine how the multiplicative noise term in Eq.~(\ref{SDE_D}) is to be evaluated. For a \textit{deterministic} differential equation
\begin{equation}
\frac{dx}{dt}=f(x,t),
\end{equation}
where $f(x,t)$ is a nonrandom and well-behaved function, we can write down the derivative on the left-hand side as the limit of a difference and then
\begin{equation}
x(t+\delta t)\approx x(t)+f(x(t),t)\delta t.
\end{equation}
In the last term, instead of the value of the function $f$ at time $t$, we could have taken its value at time $t+\delta t$, or any linear combination of the two --- this does not matter in the limit $\delta t\to 0$, i.e., in this limit the equation
\begin{eqnarray}
x(t+\delta t)&\approx &x(t)+[(1-\alpha)f(x(t),t)\nonumber\\
& &+\alpha f(x(t+\delta t),t+\delta t)]\delta t\label{determDE}
\end{eqnarray}
can be used with any $\alpha$ between zero and one. Varying $\alpha$ changes the right-hand side of Eq.~(\ref{determDE}) by a negligible amount $O(\delta t^2)$. However, for \textit{stochastic} differential equations, like Eqs.~(\ref{SDE_Dconst}) and (\ref{SDE_D}), there are complications. First, $R(t)$ is not a well-behaved function. Instead of its instantaneous value, which is not well-defined, one needs to take its average over the interval $\delta t$, which depends on $\delta t$ and is $r/\sqrt{\delta t}$, where $r$ is a random number with $\langle r \rangle =0$, $\langle r^2\rangle =1$ and no correlation between different time intervals. Then, for Eq.~(\ref{SDE_Dconst}),
\begin{equation}
x(t+\delta t)\approx x(t)+r\sqrt{2D\delta t}.
\end{equation}
Second, when $D$ is variable [as in Eq.~(\ref{SDE_D})], there is an additional uncertainty, since the result depends on where $D$ is evaluated, that is, on the value of $\alpha$ in
\begin{eqnarray}
x(t+\delta t)& \approx & x(t)\nonumber\\
& &\hspace{-1.5cm}+r\sqrt{2[(1-\alpha)D(x(t))+\alpha D(x(t+\delta t))]\delta t},
\label{stochDE}
\end{eqnarray}
and this dependence now remains even in the limit $\delta t\to 0$. Indeed, let us assume \textit{for now} that $D(x)$ is a smooth function. From now until the end of this section, we will use $x(t+\delta t)$ to denote the value calculated using Eq. (\ref{stochDE}), rather than the actual value of the coordinate at time $t+\delta t$, thus, the ``$\approx$'' sign in Eq.~(\ref{stochDE}) is replaced by the ``$=$'' sign. Expanding
\begin{eqnarray}
D(x(t+\delta t))-D(x(t))& = & \frac{dD}{dx}[x(t+\delta t)-x(t)]+O(\delta t)\nonumber\\
& &\hspace{-4cm}=\frac{dD}{dx}r\sqrt{2[(1-\alpha)D(x(t))+\alpha D(x(t+\delta t))]\delta t}+O(\delta t)\nonumber\\
& &\hspace{-4cm}=\frac{dD}{dx}r\sqrt{2D(x(t))\delta t+O(\delta t^{3/2})}+O(\delta t)\nonumber\\
& &\hspace{-4cm}=\frac{dD}{dx}r\sqrt{2D(x(t))\delta t}+O(\delta t)
\label{expansion}
\end{eqnarray}
and using this in Eq.~(\ref{stochDE}), we get
\begin{eqnarray}
x(t+\delta t)-x(t)& &\nonumber\\
& &\hspace{-2.3cm}=r\sqrt{2\left[D(x(t))+\alpha(dD/dx)r\sqrt{2D(x(t))\delta t}+O(\delta t)\right]\delta t}\nonumber\\
& &\hspace{-2.3cm}=r\sqrt{2D(x(t))\delta t}\left[1+\frac{\alpha(dD/dx)r\sqrt{2D(x(t))\delta t}}{2D(x(t))}+O(\delta t)\right]\nonumber\\
& &\hspace{-2.3cm} =r\sqrt{2D(x(t))\delta t}+r^2\alpha (dD/dx)\delta t+O(\delta t^{3/2}).\label{stochexp}
\end{eqnarray}
In Eqs.~(\ref{expansion}), (\ref{stochexp}) and below, we have not specified explicitly the point at which the derivative $dD/dx$ is taken. It is easy to check that replacing the value of $dD/dx$ at the point $x(t)$ with that at the point $x(t+\delta t)$ or vice versa introduces an extra term of order $\delta t^{3/2}$ in the last line of Eq. (\ref{stochexp}) that can be neglected, so the exact point at which the derivative is taken does not matter. On the other hand, both the first and the second term in the last line of Eq.~(\ref{stochexp}) are important in the limit $\delta t\to 0$, even though they are of different orders in $\delta t$, because $\langle r\rangle =0$, but $\langle r^2\rangle =1\ne 0$. Indeed, over a fixed time interval $\Delta t$ such that $\Delta t=N\delta t$, the contribution of the first term as a function of $N$ and $\delta t$ is $\propto \sqrt{N\delta t}=\sqrt{\Delta t}$, while that of the second term is $\propto N\delta t=\Delta t$, and both terms survive when $\delta t\to 0$. In the second term, we can replace $r^2$ with its average value 1, since the contribution of the fluctuating part over the interval $\Delta t$ is $\propto \sqrt{N}\delta t=\sqrt{\Delta t\delta t}$, which is negligible when $\delta t\to 0$. This gives
\begin{equation}
x(t+\delta t)\approx x(t)+r\sqrt{2D(x(t))\delta t}+\alpha (dD/dx)\delta t.\label{drift}
\end{equation}
It is easy to show that the same result is obtained, if in Eq.~(\ref{stochDE}) the value of $D$ is taken at a single point between the end points of the jump, as in Ref.~\cite{lancon2002}, that is, Eq.~(\ref{stochDE}) is replaced with
\begin{eqnarray}
x(t+\delta t)&=& x(t)\nonumber\\
& &\hspace{-1.5cm}+r\sqrt{2\{D[x=(1-\alpha)x(t)+\alpha x(t+\delta t)]\}\delta t}.
\end{eqnarray}

The last term in Eq.~(\ref{drift}) is deterministic and produces \textit{drift} of the particles in the direction of the diffusivity gradient known as ``spurious flow'' \cite{vankampen1992}, ``spurious drift'', or ``noise-induced drift'' \cite{coffey2005}. The drift velocity is
\begin{equation}
V=\alpha\frac{dD}{dx};\label{driftV}
\end{equation}
it vanishes when $\alpha=0$. Note that the same drift can be produced by a suitably chosen external deterministic force, so it is always possible to change the value of $\alpha$ and still get the same dynamics by introducing an additional fictitious force~\cite{sokolov2010}.

\subsection{Particle fluxes and continuum diffusion equations}
\label{fluxes}
Although the drift velocity of particles is zero for $\alpha=0$ [Eq.~(\ref{driftV})], the net \textit{flux} through a fixed point $x_0$ does not vanish in this case, even without a particle concentration (or probability density) gradient. Indeed, assuming for simplicity that $r=\pm 1$ (with no influence on the final result), in a single jump the point with $x=x_0$ is crossed from the left by one half of all particles (namely, those with $r=+1$) with $x$ between $x_1$ and $x_0$, where $x_1$ obeys the equation
\begin{equation}
x_0=x_1+\sqrt{2D(x_1)\delta t},
\end{equation}
or
\begin{eqnarray}
x_0-x_1&\approx &\sqrt{2[D(x_0)-(dD/dx)(x_0-x_1)]\delta t}\nonumber\\
& &\hspace{-1.3cm}\approx \sqrt{2D(x_0)\delta t}\left(1-\frac{1}{2D(x_0)}\frac{dD}{dx}(x_0-x_1)\right),
\end{eqnarray}
with the solution
\begin{eqnarray}
x_0-x_1&\approx &\frac{\sqrt{2D(x_0)\delta t}}{1+\frac{dD/dx}{\sqrt{2D(x_0)}}\sqrt{\delta t}}\nonumber\\
&\approx &\sqrt{2D(x_0)\delta t}-(dD/dx)\delta t.
\end{eqnarray}
The number of such particles is
\begin{equation}
n_+\approx \frac{1}{2}\left(\sqrt{2D(x_0)\delta t}-(dD/dx)\delta t\right)\rho,
\end{equation}
where $\rho$ is the (uniform) particle concentration. Likewise, the same point $x_0$ is crossed from the right by one half of all particles (those with $r=-1$) with $x$ between $x_0$ and $x_2$, with
\begin{equation}
x_2-x_0\approx \sqrt{2D(x_0)\delta t}+(dD/dx)\delta t,
\end{equation}
and the number of such particles is
\begin{equation}
n_-\approx \frac{1}{2}\left(\sqrt{2D(x_0)\delta t}+(dD/dx)\delta t\right)\rho.
\end{equation}
The net flux is thus
\begin{equation}
(n_+-n_-)/\delta t=-\rho\frac{dD}{dx}.\label{flux0}
\end{equation}
When $\alpha\ne 0$, the flux due to the drift, $\rho V$, with $V$ given by Eq.~(\ref{driftV}), has to be added to Eq.~(\ref{flux0}), which gives the total flux
\begin{equation}
J=-(1-\alpha)\rho\frac{dD}{dx}.
\end{equation}
This flux is nonzero when the drift velocity is zero at $\alpha=0$, but it vanishes when $\alpha=1$. When a concentration gradient is present [$\rho=\rho(x,t)$], this adds the usual flux given by Fick's first law, $-D \frac{\partial\rho}{\partial x}$:
\begin{equation}
J=-(1-\alpha)\rho\frac{dD}{dx}-D\frac{\partial \rho}{\partial x}.\label{flux}
\end{equation}
In two special cases this expression is simplified: for $\alpha=0$,
\begin{equation}
J=-\frac{\partial (D\rho)}{\partial x},
\end{equation}
while for $\alpha=1$,
\begin{equation}
J=-D\frac{\partial \rho}{\partial x},
\end{equation}
which conforms to normal Fickian diffusion rules. The rate of the concentration change is
\begin{equation}
\frac{\partial \rho}{\partial t}=-\frac{\partial J}{\partial x}=(1-\alpha)\frac{\partial}{\partial x}\left(\rho\frac{d D}{d x}\right)+\frac{\partial}{\partial x}\left(D\frac{\partial\rho}{\partial x}\right).\label{diffuseq}
\end{equation}
In particular, for $\alpha=0$,
\begin{equation}
\frac{\partial \rho}{\partial t}=\frac{\partial^2 (D\rho)}{\partial x^2},\label{fokker}
\end{equation}
which is the standard form of the Fokker-Planck equation without the drift term \cite{vankampen1992}; for $\alpha=1$,
\begin{equation}
\frac{\partial \rho}{\partial t}=\frac{\partial}{\partial x}\left(D\frac{\partial\rho}{\partial x}\right),
\end{equation}
which is Fick's second law.

The stationary state can be obtained by setting $\partial \rho/\partial t=0$ in Eq.~(\ref{diffuseq}). This condition for no net change in the particle concentration is then
\begin{equation}
-J=(1-\alpha)\rho\frac{dD}{dx}+D\frac{d\rho}{dx}={\rm const}.\label{equicond}
\end{equation}
If there is no net flux, as, for example, when the system is confined between two reflecting walls, then the constant in Eq.~(\ref{equicond}) is zero and
\begin{equation}
(1-\alpha)\rho dD=-Dd\rho,\label{Drhoeq}
\end{equation}
which gives
\begin{equation}
\rho D^{1-\alpha}={\rm const}.\label{equil}
\end{equation}
For $\alpha=1$, the particles can be found with equal probability anywhere ($\rho={\rm const}$), but for $\alpha<1$ they are more likely to be found in regions of lower diffusivity (or higher viscosity).

\subsection{Ito, Stratonovich, and isothermal calculi}
Even though the actual solution of Eq.~(\ref{SDE_D}) depends on the value of $\alpha$ in its discretized variant [Eq.~(\ref{stochDE})], its \textit{formal} solution can always be written as
\begin{equation}
x(t)=x(0)+\int_0^t \sqrt{2D(x(t'))}R(t')dt'.\label{formal}
\end{equation}
Different $\alpha$ thus correspond to different interpretations of the formal integral in Eq.~(\ref{formal}), or different \textit{calculi}. Three cases have received special attention.

The case $\alpha=0$ corresponds to \textit{Ito calculus}  \cite{ito1951}. In this case, the stationary distribution is given by
\begin{equation}
\rho(x)D(x)={\rm const}.\label{ItoStatio}
\end{equation}
The primary attractive quality of this formulation is that its implementation is ``non-predictive": According to Eq.~(\ref{stochDE}), the value of the diffusivity $D$ is taken at the initial point of the jump only, so, as opposed to all other cases, one does not need to look ahead in time to evaluate the spatially dependent term. This property makes the Ito formulation a frequent choice for mathematical problems --- including financial modeling. But it is also physically relevant: For instance, the overdamped limit of the Langevin equation (\ref{LDiner}) corresponds to the Ito calculus in the idealized situation where the temperature varies in space, but the friction coefficient is temperature-independent and thus the same everywhere \cite{matsuo2000}. This is an inherently non-equilibrium situation from the thermodynamic point of view and so the fact that the stationary distribution does not coincide with the equilibrium Boltzmann distribution is not surprising. Another example would be a particle in a modulated quasiperiodic potential where the tops of the barriers are all at the same level, but the well depths are position-dependent~\cite{sokolov2010}. In this case, the change in diffusivity is accompanied by the change in the free energy and the Boltzmann distribution is obeyed.

The case $\alpha=1/2$ corresponds to the \textit{Stratonovich calculus} \cite{stratonovich1966}. In this case, the value of $D$ at the midpoint of the jump (or the average of the values at the endpoints) is taken. Stratonovich calculus conforms to the ``normal'' rules of calculus. For example, if $x$ obeys Eq.~(\ref{SDE_D}), then one would expect
\begin{equation}
\frac{d}{dt}x^2=2x\frac{dx}{dt}=2x\sqrt{2D(x)}R(t),
\end{equation}
where the right-hand side is interpreted according to the rules of the given calculus. This is indeed true for Stratonovich, but not for other calculi. The Stratonovich calculus arises when considering the Langevin equation (\ref{LDiner}) with the noise $R(t)$ having a \textit{finite} and space- and time-independent correlation time, in the limit when both this correlation time and the relaxation time $m/\zeta$ go to zero, but the former is kept much larger than the latter \cite{kupferman2004}.  In fact, even for the overdamped Langevin equation~(\ref{SDE_D}) (corresponding to the relaxation time equal to zero) with a finite noise correlation time the issue of interpretation of the noise term does not arise and the limit when the correlation time goes to zero corresponds unambiguously to the Stratonovich calculus~\cite{coffey2005, vankampen1992}. This situation violates the fluctuation-dissipation theorem which requires that a finite noise correlation time be accompanied by a memory in the friction force, thus giving rise to the generalized Langevin equation \cite{pottier2009} instead of the ordinary one. 
So the fact that the density in the stationary state is not uniform, but rather obeys
\begin{equation}
\rho\sqrt{D}={\rm const},\label{StratoStatio}
\end{equation}
is natural. Sokolov~\cite{sokolov2010} has also proposed a variant of his modulated quasiperiodic potential model that corresponds to the Stratonovich calculus; again, the free energy in that model is position-dependent. Since Ito and Stratonovich are the two calculi most often discussed in the literature, the choice between different calculi has often been referred to as the ``Ito-Stratonovich dilemma''~\cite{vankampen1992}.

But besides these two cases, the case $\alpha=1$ is also of special interest, because in that case Eq.~(\ref{equil}) becomes
\begin{equation}
\rho={\rm const},\label{IsoStatio}
\end{equation}
which corresponds to the equilibrium Boltzmann distribution in the ``purest'' case when it is only the diffusivity that varies in space, but the particle free energy and the temperature are the same everywhere. There is no universally accepted name for the $\alpha=1$ calculus, with names such as H\"anggi-Klimontovich \cite{sokolov2010}, kinetic \cite{sokolov2010, klimontovich1990} and isothermal \cite{lancon2002} appearing in the literature. We will use the latter in what follows, since it emphasizes the fact that this calculus corresponds to the case when the system is kept at a constant temperature, as opposed to the Ito case which can be achieved by varying the temperature across the system (still assuming the situation where the free energy would be constant for $T={\rm const}$).

\subsection{Discontinuous $D(x)$}

So far, we have assumed that $D(x)$ varies smoothly in space. First of all, this allowed us to do the expansion in Eq.~(\ref{expansion}). Even more importantly, we note that if there is a jump in the diffusivity, then even the solution of Eq.~(\ref{SDE_D}) is not well-defined for $\alpha\ne 0$. Consider $\alpha=1$ and let us assume that
\begin{equation}
D(x)=\left\{
\begin{array}{l l}
D_L, & \ x<0,\\
D_R, & \ x>0,\\
\end{array}
\right.
\end{equation}
with $D_L>D_R$. Let the particle coordinate at time $t$ be $x_0<0$ and suppose that the time step $\delta t$ and the random number $r$ chosen at the following step are such that $x_0+\sqrt{2D_L}r\delta t>0$, but $x_0+\sqrt{2D_R}r\delta t<0$. Note that the jump size depends on the final position at time $t+\delta t$. If we assume that the particle ends up to the right of the interface, then the jump size should be $\sqrt{2D_R}r\delta t$, but then the particle coordinate after the jump is $x_0+\sqrt{2D_R}r\delta t<0$, so we come to a contradiction. Conversely, if we assume that the particle ends up on the left side, then the jump size should be $\sqrt{2D_L}r\delta t$, but $x_0+\sqrt{2D_L}r\delta t>0$, so there is again a contradiction. Thus at this point the particle behavior is undefined. On the other hand, for $x_0>0$ it is possible that both assumptions lead to valid results, so there is ambiguity.
Given that in principle $|x_0|$ can be arbitrarily close to zero, this will always be a potential problem, no matter how small $\delta t$ is. In fact, this problem exists whenever the jump size depends not just on the initial position, that is, for any $\alpha\ne 0$. 
Thus, in order to be able to use Eq.~(\ref{SDE_D}) [discretizing it as in Eq.~(\ref{stochDE})] for calculi other than Ito (and even for the very concept of different calculi to be meaningful), we need to assume that the function $D(x)$ is smooth everywhere. 
In that case, what we mean by a ``sharp'' interface and a ``jump'' in diffusivity is that actually the diffusivity is smooth and thus the interface between two liquids has a finite width, but the interface is narrow compared to all other length scales (in our simple problem, there is only one such length scale, the size of the system). 
Physically, the smoothness of $D$ means that it changes little on the length scale of the typical interatomic distance. This is very likely to be true in reality, and even if the liquids are completely immiscible and the interface is atomically sharp, but the diffusing particle is much larger than the interatomic distance, the properties of the system are effectively averaged out over the particle size and the condition is satisfied.
In this case, Eq.~(\ref{SDE_D}) is still meaningful for all calculi and in principle can be solved numerically by choosing a very small time step when iterating Eq.~(\ref{stochDE}), so that typical particle jumps in the simulation are smaller than the width of the interface. However, this would be a waste of computational resources; but choosing a larger step would lead to the indeterminacy problem described above. This is where LMC methods are helpful, as we will see below.

In the remainder of this paper, we will concentrate on this situation of a sharp, but not infinitely sharp interface. 
For simplicity, we will still use notation that may seem to imply that the interface is infinitely sharp, but will keep in mind that in reality it is not. 
So, for instance, $\rho(+0,t)$ denotes the particle concentration immediately to the right of the interface, but still far enough from it that the diffusivity has already reached its asymptotic value $D_R$; likewise, $D(+0)\equiv D_R$. 
Bearing this in mind, Eq.~(\ref{equil}) for the stationary distribution and its particular cases, Eqs.~(\ref{ItoStatio}), (\ref{StratoStatio}), and (\ref{IsoStatio}), should remain valid in our situation. 
Moreover, we can show that these relations are valid \textit{at all times}, even in the transient state, \textit{in the vicinity of the interface}, that is, at all times
\begin{equation}
\rho(-0,t)[D(-0)]^{1-\alpha}=\rho(+0,t)[D(+0)]^{1-\alpha}.\label{jump}
\end{equation}
Indeed, with the exception of the case $\alpha=1$, the first term on the right-hand side of Eq.~(\ref{flux}) for the particle flux is large at the interface and in fact infinite in the limit of an infinitely sharp interface. However, the flux itself, even though it can be nonzero when the stationary state is not reached, is expected to be finite. Therefore, the second term on the right-hand side of Eq.~(\ref{flux}) should be large (infinite in the limit) as well, and the flux $J$ can be neglected compared to either of these terms. Dropping $J$ in Eq.~(\ref{flux}) produces the stationary-state equation~(\ref{Drhoeq}), from which Eq.~(\ref{jump}) follows immediately. In the special case $\alpha=1$, the first term on the right-hand side of Eq.~(\ref{flux}) is zero and therefore the second term is finite, thus there is no jump in $\rho$, which is likewise consistent with Eq.~(\ref{jump}).
Condition (\ref{jump}) describes the concentration jump across the interface that forms immediately and always exists, if $\alpha\ne 1$. 
The full solution of the diffusion problem for the system consisting of two uniform regions separated by a ``sharp'' interface (where ``sharp'' is understood as above) can be obtained by solving two uniform diffusion problems in the two regions,
\begin{equation}
\frac{\partial\rho(x,t)}{\partial t}=D_{L,R}\frac{\partial^2 \rho(x,t)}{\partial x^2},\label{diffpieces}
\end{equation}
with two additional matching conditions at the interface: Eq.~(\ref{jump}) and the flux continuity condition,
\begin{equation}
D_L\left.\frac{\partial\rho(x,t)}{\partial x}\right|_{x=-0}=D_R\left.\frac{\partial\rho(x,t)}{\partial x}\right|_{x=+0}.\label{fluxcont}
\end{equation}
In addition, of course, there have to be boundary conditions at the walls, for example, the no-flux condition
\begin{equation}
\left.\frac{\partial\rho(x,t)}{\partial x}\right|_{x=x_b}=0
\end{equation}
for a reflecting boundary or
\begin{equation}
\rho(x_b,t)=0
\end{equation}
for an absorbing boundary, where $x_b$ is the coordinate of the boundary.

\section{Molecular Dynamics Simulation Approaches}

In this section and the next one, the various simulation approaches used to explore diffusion in an inhomogeneous medium are presented. 
We start by describing two well-known molecular dynamics (MD) approaches, namely, Langevin dynamics (LD) and Brownian dynamics (BD).

MD is a simulation approach based on solving equations of motions for particles comprising the system. In the most straightforward, but very computationally intensive approach, both the probe particle(s) whose diffusion is studied and the fluid particles are included explicitly and Newton's Second Law equations
\begin{equation}
m_i \ddot{\vec{x}}_i=\vec{F}_i
\end{equation}
are solved numerically, where $m_i$ and $\vec{x}_i$ are the mass and the position of the $i$th particle and $\vec{F}_i$ is the sum of all forces acting on it. The computational expense can be reduced by considering the fluid particles implicitly, as is done in the LD and BD approaches.

\subsection{Langevin Dynamics}
In the LD approach (\cite{schlick2010}, Sec. 14.4), only the equation of motion of the diffusing particle is considered explicitly and the action of the fluid particles on it is replaced by the friction force and a random force, and the resulting equation of motion is given in 1D by Eq.~(\ref{LDiner}) with $U=0$:
\begin{equation}
m \ddot{x} = - \zeta \dot{x} + \sqrt{2 kT \zeta} R(t).
\end{equation}
Dividing by $\zeta$ and assuming that $\zeta$ is a function of $x$, we obtain
\begin{equation}
\frac{m}{\zeta(x)} \ddot{x} = - \dot{x} + \sqrt{\frac{2 kT}{\zeta(x)}} R(t),
\end{equation}
or
\begin{equation}
\frac{mD(x)}{kT} \ddot{x} = - \dot{x} + \sqrt{2 D(x)} R(t).
\label{LD}
\end{equation}
Unlike in the case of the first-order Eq.~(\ref{SDE_D}), there is no ambiguity due to different interpretations of the random term even when $\zeta$ (or $D$) is space-dependent, with all interpretations and all ``correct'' numerical methods giving the same solution in the limit when the time step $\delta t\to 0$. This is because, unlike in the case of Eq.~(\ref{SDE_D}), the particle velocity does not diverge, thus when the equation is discretized, the displacement during a single step is much smaller [$O(\delta t)$ instead of $O(\delta t^{1/2})$], so in the limit $\delta t\to 0$ it no longer matters whether the position at the beginning or at the end of the step is used when calculating $D$~\cite{matsuo2000}. In practice, some variation of the Verlet method (\cite{schlick2010}, Sec. 13.4) is usually used to solve Eq.~(\ref{LD}) numerically. By construction, Eq.~(\ref{LD}) obeys the fluctuation-dissipation theorem, thus in the case $T={\rm const}$ the stationary distribution should coincide with the Boltzmann distribution $\rho={\rm const}$. Since this fact does not depend on the value of the parameters entering the equation, it should remain valid in the limit $m\to 0$ when the inertial term vanishes, and thus this overdamped limit should correspond to the isothermal interpretation of Eq.~(\ref{SDE_D}). In fact, this was proved in Ref.~\cite{sancho1982}. 
However, in order to reproduce this result numerically, it is important to retain the inertial term decreasing its coefficient, rather than simply dropping it, and to make sure that the time step is always much smaller than the relaxation time, $m/\zeta$. Remarkably, unlike Eq.~(\ref{SDE_D}) with $\alpha\ne 0$, Eq.~(\ref{LD}) can still be solved straightforwardly in the case of a sharp interface, as we will see in Sec.~VI.

\subsection{Brownian Dynamics}
The motion of microscopic particles in a liquid is nearly always overdamped, with the time of relaxation to the terminal velocity, $m/\zeta$, usually much smaller than any other relevant time scale. As mentioned above, the time step of the LD algorithm should be much smaller than even this extremely small time. Even if the particle mass $m$ is increased artificially to increase the relaxation time and thus the maximum allowed time step, this time step may still end up being too small and as a result, the simulation will be inefficient. A seemingly straightforward approach is eliminating the left-hand side of Eq.~(\ref{LD}) (the inertial term) altogether and solving the resulting first-order equation. This equation coincides with Eq.~(\ref{SDE_D}), so the discussion of the ambiguity of the stochastic term in Sec.~\ref{sec:theory} applies. Normally in MD simulations, to solve Eq.~(\ref{SDE_D}) numerically the simplest Euler algorithm (often referred to as Euler-Maruyama in the stochastic case \cite{kloeden1992}) is used, in which, in order to obtain the particle position at time $t+\delta t$, the right-hand side of the equation is evaluated at time $t$. This approach, called Brownian dynamics, obviously corresponds to the Ito interpretation of the stochastic differential equation (\ref{SDE_D}). Thus, surprisingly, LD (even in the overdamped limit) and BD will give different results when the friction coefficient $\zeta$ varies in space, as they correspond to different calculi. Just as LD, BD can be used in the case of a sharp interface without any complications, since these complications only arise for $\alpha\ne 0$ when solving Eq.~(\ref{SDE_D}). 
Some difficulties arise when the random number $r$ is chosen to be discrete (e.g., $\pm 1$), as described in Sec.~\ref{sec:MCIIvsBD}, so using $r$ with a continuous distribution is preferable.

\section{Monte Carlo Approaches}
The main goal of this paper is to develop and test LMC algorithms for simulating a random walker at a viscosity interface corresponding to different calculi [$\alpha$ in Eq.~(\ref{stochDE})], including the popular Ito, Stratonovich, and isothermal calculi. There are several reasons to develop such algorithms. First, LMC algorithms are popular for studying diffusion problems
\cite{lopez2002,bernstein2005,saxton1996,plischke99,sahimi02,sinnoKMC09,keller02,guo96,mercier00,mercier01macro,mercier01noncond,gauthier02,gauthier03,casault07,kosmidis,torres08,langowski} 
and in some of these problems studying the case when the effective viscosity is inhomogeneous in space (e.g., different degrees of crowding in different parts of the system) can be of interest (see, e.g., Ref.~\cite{bernstein2005}). Second, LMC algorithms can offer advantages compared to MD (or, in general, solving numerically the Langevin equation) specifically in the case of a sharp interface. As we have seen, sharp interfaces are algorithmically problematic when solving the overdamped Langevin equation, except in the case $\alpha=0$ corresponding to BD. While the $\alpha=1$ (isothermal) case can be treated by LD, this may require a very small time step for the solution to be reliable. All other cases are in principle equivalent to Ito plus an additional drift force, but that drift force would be infinite at the interface and zero elsewhere, which would cause obvious numerical problems. LMC algorithms, on the other hand, are straightforward in all of these cases, as we will demonstrate.

Two different approaches will be taken. In the first method, changes in viscosity values are introduced by altering the probability $p_s$ that a particle does not jump during a MC jump attempt: the particle has a higher chance of ``staying put'' in the high viscosity region. In this approach, the jump lengths on the low and high viscosity sides are the same. Conversely, in the second method, the change in viscosity across the interface is included by altering the jump length (i.e., the mesh step of the lattice): The particle jumps shorter distances when the viscosity is higher. Here, the probability of not jumping is constant (except perhaps for sites adjacent to the interface). In both methods the time step is the same in both regions and stays constant throughout the simulation. This is essential for so-called numerically exact variants of LMC methods \cite{majid1984,guo96,mercier1999,gauthier02,casault07,gauthier2008,torres08} and is also very helpful when several particles are simulated at once.

Both LMC methods are derived based on three requirements: (1) the diffusion should be unbiased away from the interface, (2) the diffusion rates should be correct in both regions, and (3) the probabilities of the moves at the interface should ensure that the stationary distribution given by Eq.~(\ref{equil}) is reproduced. Only the last of these requirements depends on the calculus (the value of $\alpha$), and so this value will only affect the rules of the algorithms at the interface, that is, in the only place where the viscosity gradient is present, as expected.

When deriving both algorithms, the following consideration is used. Suppose the particle moves on a lattice with a constant mesh step $a$ and the time step is $\tau$. At every \textit{successful} step (i.e., ignoring the time steps where the particle stays put), the particle moves left or right with an equal probability. The mean-square displacement after $N$ successful steps is
\begin{equation}
\langle \Delta x^2 \rangle = Na^2.
\end{equation}
Since in time $\Delta t$ there are $\Delta t/\tau$ total steps and thus on average $\Delta t(1-p_s)/\tau$ successful steps,
\begin{equation}
\langle \Delta x^2 \rangle = \frac{(1-p_s)a^2}{\tau}\Delta t,
\end{equation}
and comparing this to Eq.~(\ref{D}),
\begin{equation}
D=\frac{(1-p_s)a^2}{2\tau}.\label{D_LMC}
\end{equation}

\subsection{MC I: Variable $p_s$, fixed jump length $a$}
\label{MCI}

\subsubsection{The basic algorithm}
\begin{figure}[h!]
 	\centering
	\includegraphics[width=0.45\textwidth]{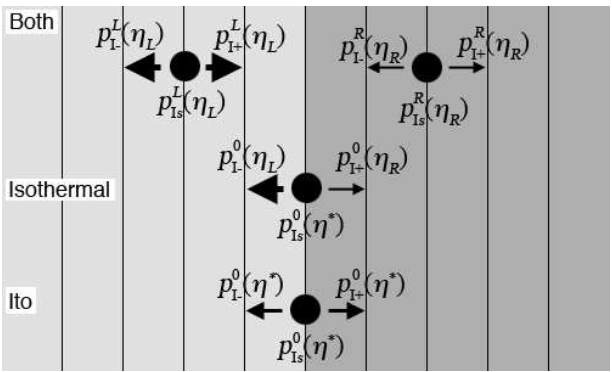}
	\caption{A schematic of the MC I algorithm. Different shades of gray depict regions of different viscosity, with the darker region being more viscous. Vertical lines correspond to the locations of 1D lattice sites. The distance between sites is the same in both regions. Arrows show transitions between sites: the thicker the arrow, the higher the transition probability. The notation for the probabilities corresponds to that used in the text. The top part illustrates the behavior of a particle away from the interface, which is independent of the calculus, while the middle and bottom parts are for a particle at the interface site in the two particular cases of the isothermal and Ito calculi, respectively.}
	\label{fig:MCI}
\end{figure}

In the first Monte Carlo approach, all ``successful" jumps have the same length and the dependence on viscosity is introduced by altering the probability of remaining still during a time step (Fig.~\ref{fig:MCI}).

Given the diffusion coefficient $D$ and the time step $\tau$, from Eq.~(\ref{D_LMC}) the probabilities to move right and left should be
\begin{equation}
p_{+}=p_{-}=\frac{1-p_s}{2}=\frac{D\tau}{a^2},\label{ppm1}
\end{equation}
and the probability to stay put,
\begin{equation}
p_s=1-\frac{2D\tau}{a^2}.
\end{equation}
Obviously, $\tau$ and $a$ need to be chosen so that all of these probabilities are between zero and one for the diffusivities both to the left ($D=D_L$) and to the right ($D=D_R$) of the interface. Since $D$ is inversely proportional to the fluid viscosity $\eta$ [see Eq.~(\ref{diffzeta})], it follows from Eq.~(\ref{ppm1}) that $p_{+}$ and $p_{-}$ should be inversely proportional to $\eta$, or, introducing the viscosity $\eta^{\I}_0$ at which these probabilities are equal to 1/2 and correspondingly $p_s=0$,
\begin{eqnarray}
p_{\I +}^{L,R}&=&p_{\I -}^{L,R}=\frac{\eta^{\I}_0}{2\eta_{L,R}},\label{ppmLR1}\\
p_{\I s}^{L,R}&=&1-\frac{\eta^{\I}_0}{\eta_{L,R}},\label{psLR1}
\end{eqnarray}
where sub- and superscripts $L$ and $R$ denote the left and right side of the viscosity interface, respectively, and ``I'' shows that these parameters correspond to the MC I algorithm. The same probabilities were used previously in Refs.~\cite{hickey2006,kingsburry2009}. The parameter $\eta^{\I}_0$ should not be larger than the smaller of the two viscosities $\eta_L$ and $\eta_R$ to ensure that both $p^L_{\I s}$ and $p^R_{\I s}$ are non-negative. In fact, even the situation where $\eta^{\I}_0$ is equal to or only slightly smaller than either $\eta_L$ or $\eta_R$ should be avoided, since some simulation artifacts, such as spurious oscillations both in time and in space, are possible when the waiting time is zero~\cite{chubynsky2012}, or, more generally, the variance of the time between successful steps is small.  
On the other hand, from Eqs.~(\ref{diffzeta}) and (\ref{ppm1}),
\begin{equation}
\eta^{\I}_0=\frac{2kT\tau_{\I}}{\kappa a_{\I}^2},\label{eta0I}
\end{equation}
so this limitation on the allowed values of $\eta^{\I}_0$ restricts possible choices of $\tau_{\I}$ and $a_{\I}$.
The fact that the probabilities of the moves are different in the two parts of the system is depicted schematically in the top part of Fig.~\ref{fig:MCI}.

Equations~(\ref{ppmLR1}) and (\ref{psLR1}) should be used for lattice sites away from the interface. 
They ensure that the first two of the three requirements mentioned above (no bias and the correct diffusivities) are satisfied. 
The third condition, the correct ratio of the particle concentrations on the two sides of the interface, should be achieved by choosing a special set of probabilities for sites near the interface. 
It is convenient to choose the lattice so that the interface is on a lattice site. 
Let us denote this site 0, and the sites immediately to the left and to the right of it, $-1$ and $+1$, respectively. 
Since we have a single condition to satisfy for several probabilities, we can make an arbitrary choice that the probabilities of moves out of sites $-1$ and $+1$ are still given by Eqs.~(\ref{ppmLR1}) and (\ref{psLR1}) and that only the two probabilities for the moves out of site 0 (plus the rest probability at site 0) are special. 
Based on Eq.~(\ref{equil}), the condition that needs to be satisfied is that
\begin{equation}
\frac{n_1}{n_{-1}}=\left(\frac{\eta_R}{\eta_L}\right)^{1-\alpha},
\end{equation}
where $n_i$ is the average number of particles at site $i$ in the stationary state. Since from detailed balance,
\begin{equation}
\frac{n_1}{n_0}=\frac{p^0_{\I +}}{p^R_{\I -}}
\end{equation}
and
\begin{equation}
\frac{n_0}{n_{-1}}=\frac{p^L_{\I +}}{p^0_{\I -}},
\end{equation}
where $p^0_{\I +}$ and $p^0_{\I -}$ are the probabilities of the moves from site 0 to the right and to the left, respectively, we get
\begin{equation}
\frac{n_1}{n_{-1}}=\frac{p^0_{\I +}}{p^{0}_{\I -}}\frac{p^L_{\I +}}{p^R_{\I -}}=\frac{p^0_{\I +}}{p^{0}_{\I -}}\frac{\eta_R}{\eta_L}=\left(\frac{\eta_R}{\eta_L}\right)^{1-\alpha},
\end{equation}
or
\begin{equation}
\frac{p^0_{\I -}}{p^{0}_{\I +}}=\left(\frac{\eta_R}{\eta_L}\right)^{\alpha}.\label{ratio1}
\end{equation}
This means that for $\eta_R\ne\eta_L$ and $\alpha\ne 0$ there is a bias at the interface ($p^0_{\I -}\ne p^0_{\I +}$).

Equation~(\ref{ratio1}) defines the ratio $p^0_{\I -}/p^0_{\I +}$, but not each of these probabilities separately. 
In the isothermal case ($\alpha=1$), $\rho={\rm const}$ in the stationary state and thus $n_1=n_{-1}$; it is reasonable to require that $n_0$ is the same as $n_{-1}$ and $n_1$, and this gives unambiguously
\begin{eqnarray}
p^{0}_{\I -}(\alpha=1)&=&p^L_{\I +}=\frac{\eta^{\I}_0}{2\eta_L},\\
p^{0}_{\I +}(\alpha=1)&=&p^R_{\I -}=\frac{\eta^{\I}_0}{2\eta_R}.
\end{eqnarray}
The probability of staying put is then
\begin{equation}
p^{0}_{\I s}=1-p^{0}_{\I -}-p^{0}_{\I +}=1-\frac{\eta^{\I}_0}{\eta^*},\label{ps1int}
\end{equation}
where the reduced viscosity
\begin{equation}
\eta^*=\frac{2\eta_L\eta_R}{\eta_L+\eta_R}
\end{equation}
is the harmonic mean of the two viscosities. Since it makes sense for the rate of moving away from the interface to always be related to some sort of average between the two viscosities, we can require, somewhat arbitrarily, that the probability of staying put be given by Eq.~(\ref{ps1int}) for all values of $\alpha$ and use this as an additional condition together with Eq.~(\ref{ratio1}) to obtain
\begin{eqnarray}
p^{0}_{\I -}&=&\frac{\eta^{\I}_0}{\eta^*}\frac{\eta_R^{\alpha}}{\eta_L^{\alpha}+\eta_R^{\alpha}},\label{p0m1}\\
p^{0}_{\I +}&=&\frac{\eta^{\I}_0}{\eta^*}\frac{\eta_L^{\alpha}}{\eta_L^{\alpha}+\eta_R^{\alpha}}.\label{p0p1}
\end{eqnarray}
In fact, we have checked for the Ito calculus that the dynamics are insensitive to the choice of $p^0_{\I s}$ provided that $p^{0}_{\I +}=p^{0}_{\I -}$, which in this case follows directly from Eq.~(\ref{ratio1}). Results essentially indistinguishable from those generated by the above formulation were obtained in simulations with $p^0_{\I s}=0$ and $p^0_{\I \pm}=0.5$.

Equations~(\ref{p0m1}) and (\ref{p0p1}) have simple interpretations in two cases. In the isothermal case ($\alpha=1$), $p^0_{\I -}=p^L_{\I \pm}$ and $p^0_{\I +}=p^R_{\I \pm}$. In other words, the probability of the move to the left from the interface site (the move that is within the left region) is the same as the probability of any other move (from a non-interface site) within the left region, and the same is true for the right region. This property is illustrated in the middle part of Fig.~\ref{fig:MCI}. It ensures explicitly the absense of a net flux through any surface when the particle concentration is the same everywhere, the key property of the $\alpha=1$ case mentioned in Sec.~\ref{sec:theory}, since for every surface drawn between two sites there will be as many particles moving from left to right as from right to left. In the Ito case ($\alpha=0$), the probabilities of moves to the left and to the right from the interface site are equal (the bottom part of Fig.~\ref{fig:MCI}). This is consistent with the description of the Ito formulation in Secs.~\ref{sec:intro} and \ref{sec:theory} as the case where any bias at the interface is absent.

\subsubsection{Variants with all move attempts successful}
\label{MCIallsucc}
When the ratio of the two viscosities, $\eta_L$ and $\eta_R$, is very large or very small, then, since $\eta^{\I}_0$ should be chosen at least as small as the smaller of the two, the probability of moving in the more viscous region will be very low, according to Eq.~(\ref{ppmLR1}). This makes the algorithm very inefficient in this case, since many random numbers need to be generated per single successful move. However, one may note that the number of steps until the next successful move is exponentially distributed, with the probability that this number is $n$ being
\begin{equation}
p_{\I n}=(1-p_{\I s})p_{\I s}^{n-1},\label{ndistr}
\end{equation}
where $p_{\I s}$ is either $p^L_{\I s}$, $p^R_{\I s}$, or $p^0_{\I s}$, depending on the location of the particle. Instead of waiting for many steps until a successful move, one can move at every step, but advance the clock by $n\tau_{\I}$ instead of $\tau_{\I}$, generating $n$ from the distribution (\ref{ndistr}). The resulting procedure is exactly equivalent to the original one. Strictly speaking, in this procedure the time step is variable, which in some cases is undesirable, as mentioned above; however, it is still always a multiple of the basic step $\tau_{\I}$. It is also possible to always advance the clock by the same amount $t_{\I w}$, equal to the average waiting time between successful steps,
\begin{equation}
t_{\I w}=\frac{\tau_{\I}}{1-p_{\I s}}=\frac{\eta}{\eta^{\I}_0}\tau_{\I},\label{allsucc}
\end{equation}
where $\eta$ is either $\eta_L$, $\eta_R$, or $\eta^*$.
While not strictly equivalent to the algorithm given by Eqs.~(\ref{ppmLR1}) and (\ref{psLR1}), such an algorithm should ultimately give the same results in the limit $a,\tau\to 0$. The advantage is that the need to draw random numbers from the distribution (\ref{ndistr}) is avoided. However, the time step will no longer be a multiple of $\tau$. A potentially more serious problem is simulation artifacts that may appear when the waiting time is constant, as mentioned above. Regardless of the utility of this algorithm for actual simulations, it is useful in the analysis of the diffusion problem, as we discuss below.

In both variants of the algorithm (with random and deterministic clock advancement), at each step the moves to the left and to the right are equiprobable, \textit{except} when the particle is right at the viscosity interface the probabilities are
\begin{eqnarray}
p'^0_{\I -}&=&\frac{\eta_R^{\alpha}}{\eta_L^{\alpha}+\eta_R^{\alpha}},\label{pminussucc}\\
p'^0_{\I +}&=&\frac{\eta_L^{\alpha}}{\eta_L^{\alpha}+\eta_R^{\alpha}}.\label{pplussucc}
\end{eqnarray}
These probabilities are such that their ratio $p'^0_{\I +}/p'^0_{\I -}$ is the same as $p^0_{\I +}/p^0_{\I -}$, but their sum is unity, so the rest probability is zero.

Note that in the algorithm presented here the particle moves to a neighboring site at every step, that is, the particle motion is simply an ordinary random walk on a 1D lattice, and it is unbiased everywhere, except perhaps at the interface. 
(We ignore for the moment the fact that each step takes a different amount of time in different regions.)
In fact, in the special case of the Ito calculus the random walk is unbiased at the interface as well, since in that case Eqs.~(\ref{pminussucc}) and (\ref{pplussucc}) give $p'^0_{\I -}=p'^0_{\I +}$. 
In general, the trajectory of the particle can be thought of as consisting of segments of an unbiased random walk terminating at the interface; every time the interface is reached, the particle chooses to continue on the left side with the probability $p'^0_{\I -}$ or on the right side with the probability $p'^0_{\I +}$. 
Each such choice is independent of all of the previous choices and of whether the particle has reached the interface from the left or from the right. We will use this consideration in Sec. VIB and in Appendix B to derive some properties of the system.

\subsection{MC II: Fixed $p_s$, variable jump length}
\label{sec:MCII}
\begin{figure}[h!]
 	\centering
	\includegraphics[width=0.45\textwidth]{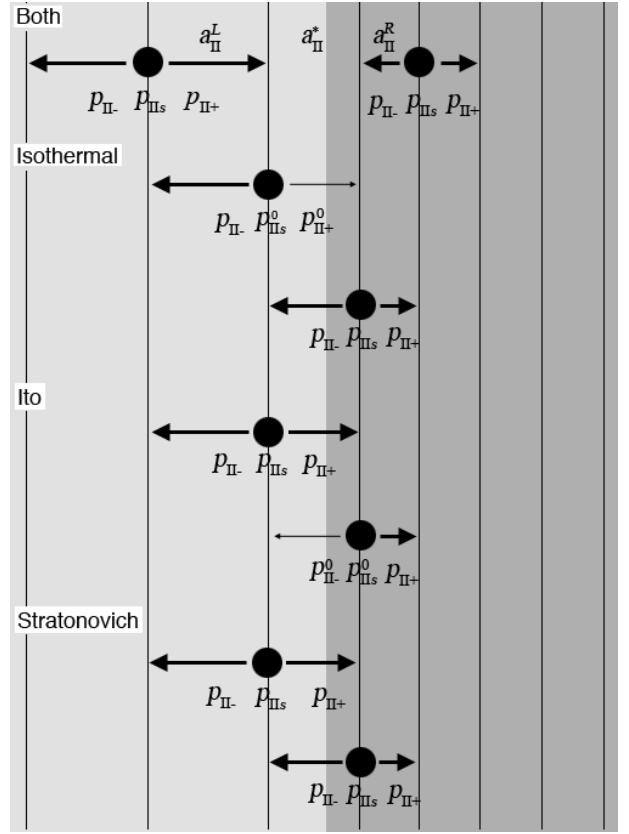}
	\caption{A schematic of the MC II algorithm. Different shades of gray depict regions of different viscosity, with the darker region being more viscous. Vertical lines correspond to the locations of 1D lattice sites. The distance between sites is smaller in the region of higher viscosity. Arrows show transitions between sites: the thicker the arrow, the higher the transition probability. Notation for the probabilities corresponds to that used in the text. The top part illustrates the behavior of a particle away from the interface, which is independent of the calculus, while the rest of the figure is for particles on sites adjacent to the interface on both sides in the cases of the isothermal, Ito, and Stratonovich calculi.}
	\label{fig:MCII}
\end{figure}

In the second MC approach, the probability of remaining still, $p_{\II s}$, is independent of the viscosity and instead the length of a successful jump is dependent on the viscosity. To keep the particle on-lattice, we introduce a mesh with a different step in the two regions (Fig.~\ref{fig:MCII}). From Eqs.~(\ref{D_LMC}) and (\ref{diffzeta}), when $p_{\II s}$ and $\tau_{\II}$ are fixed, the mesh step $a\propto\sqrt{D}\propto 1/\sqrt{\eta}$, so the mesh steps on the left and right sides are, respectively,
\begin{eqnarray}
a^L_{\II}&=&a_{\II 0}\sqrt{\frac{\eta^{\II}_0}{\eta_L}},\label{al}\\
a^R_{\II}&=&a_{\II 0}\sqrt{\frac{\eta^{\II}_0}{\eta_R}},\label{ar}
\end{eqnarray}
where $a_{\II 0}$ and $\eta^{\II}_0$ are constants and
\begin{equation}
\eta^{\II}_0=\frac{2kT\tau_{\rm II}}{\kappa a_{\II 0}^2(1-p_{\II s})}.\label{eta0II}
\end{equation}
While it is possible to put the interface at one of the lattice sites, as in MC I, we now place it between two sites to illustrate the possibility of this choice. The distance between the two sites closest to the interface (one on each side) is chosen to be the average of the mesh steps on the two sides, $a^{*}_{\II}=(a^L_{\II}+a^R_{\II})/2$, although the exact value does not matter in the limit $a^L_{\II},a^R_{\II}\to 0$.

For all sites not adjacent to the interface, on both sides, the probabilities of the moves are the same, with
\begin{equation}
p^{L,R}_{\II +}=p^{L,R}_{\II -}=(1-p_{\II s})/2\label{ppm2}
\end{equation}
and the probability of remaining still, $p_{\II s}$, is the same at all these sites, as well. The choice of $p_{\II s}$ is in principle arbitrary. While $p_{\II s}=0$ seems the most efficient choice at first glance, since the particle would move at every step of the algorithm, this, as mentioned, can introduce artifacts and, in fact, we have recently shown~\cite{chubynsky2012} that $p_{\II s}=2/3$ is optimal from the point of view of accuracy of the algorithm and may even be more efficient than $p_{\II s}=0$, since a coarser mesh can be used to achieve the same accuracy. For this reason, we choose $p_{\II s}=2/3$ in this work.

As in MC I, the probabilities of the moves need to be modified for sites adjacent to the interface. A reasonable choice is to allow the probabilities of the two moves crossing the interface to change but keep the probabilities of the two other moves leaving these sites that do not cross the interface the same as given by Eq.~(\ref{ppm2}). We use the following notation for the probabilities of the moves that are allowed to change: $p^0_{\II +}$ denotes the probability of the move to the right from the site left of the boundary and $p^0_{\II -}$ is the probability of the move to the left from the site right of the boundary. This notation is similar to that we used for MC I, but it has a somewhat different meaning: in MC I these were the probabilities of the moves from the same interface site, but here these moves cross the interface starting from different sites.

The detailed balance condition is
\begin{equation}
\frac{n_R}{n_L}=\frac{p^0_{\II +}}{p^0_{\II -}},
\end{equation}
where $n_L$ ($n_R$) is the average number of particles per site to the left (right) of the interface in the stationary state. To relate this condition to the stationary distribution of Eq.~(\ref{equil}), we need to keep in mind that the density of sites (their number per unit length) is different in the two regions. The particle concentration $\rho$ is proportional to both the number of particles per site and the site density; the latter is inversely proportional to the mesh step.  Therefore, using Eqs.~(\ref{al}) and (\ref{ar}) for the mesh steps,
\begin{equation}
\frac{\rho_R}{\rho_L}=\frac{n_R/a^R_{\II}}{n_L/a^L_{\II}}=\frac{p^0_{\II +}}{p^0_{\II -}}\sqrt{\frac{\eta_R}{\eta_L}}=\left(\frac{\eta_R}{\eta_L}\right)^{1-\alpha},
\end{equation}
or
\begin{equation}
\frac{p^0_{\II +}}{p^0_{\II -}}=\left(\frac{\eta_R}{\eta_L}\right)^{1/2-\alpha}.\label{ratio2}
\end{equation}
Again, as in MC I [cf. Eq.~(\ref{ratio1})], we have one equation for two probabilities. As an additional condition, we choose
\begin{equation}
{\rm max}(p^0_{\II +},p^0_{\II -})=p^{L,R}_{\II\pm},
\end{equation}
which, together with Eq.~(\ref{ratio2}), gives
\begin{eqnarray}
p^0_{\II +}&=&p^{L,R}_{\II\pm}{\rm min}\left[1,\left(\frac{\eta_R}{\eta_L}\right)^{1/2-\alpha}\right],\label{p0p}\\
p^0_{\II -}&=&p^{L,R}_{\II\pm}{\rm min}\left[1,\left(\frac{\eta_L}{\eta_R}\right)^{1/2-\alpha}\right]\label{p0n}.
\end{eqnarray}
This is a simple choice that guarantees that both $p^0_{\II +}$ and $p^0_{\II -}$ are always between zero and one for any value of $\alpha$ and any choice of $p_{\II s}$. Note that always either $p^0_{\II +}=p^{L,R}_{\II\pm}$ or $p^0_{\II -}=p^{L,R}_{\II\pm}$, so only one of these probabilities (at most) gets modified compared to the bulk value. The corresponding probability of staying put is changed accordingly. In the special case $\alpha=1/2$ (Stratonovich calculus), $p^0_{\II +}=p^0_{\II -}=p^{L,R}_{\II\pm}$, so all probabilities of all moves are the same (Fig. \ref{fig:MCII}). While this means that for $\alpha=1/2$ there is no flux across the interface when $n_R=n_L$, this is not the same as $\rho_R=\rho_L$ (a uniform particle distribution).

We note that a similar, but off-lattice approach was successfully used recently by one of the authors~\cite{dehaan2011} to treat a problem corresponding to the $\alpha=1$ case.

\subsection{Comparison between MC II and BD}
\label{sec:MCIIvsBD}
It is instructive to compare the MC II algorithm described here in the Ito case ($\alpha=0$) to the BD algorithm, a single step of which is given by Eq.~(\ref{stochDE}) with $\alpha=0$, specifically its variant with the random number $r=\pm 1$. Similarities are obvious: in both cases the time step is fixed and the jump size depends on what region the particle is in, being proportional to $\sqrt{D}$ (or inversely proportional to $\sqrt{\eta}$). However, what happens at the interface is different for these two algorithms. In MC II, in order to stay on-lattice, the lengths of the jumps across the interface are chosen to be the same in both directions, but in BD the jump length depends on what region the jump originates in and thus is different for the left-to-right jumps and the right-to-left jumps. This difference is crucial and makes it possible to achieve the correct particle concentration jump across the interface in the stationary state while keeping the probabilities of all jumps the same (effectively equal to $1/2$, since there is no bias in BD and no staying at rest), whereas in MC II obtaining this concentration jump requires setting $p^0_{+}\ne p^0_{-}$. This is actually a rather subtle issue. Indeed, in the stationary state, the left-to-right and the right-to-left fluxes across the interface should be equal. But then one could argue for the BD algorithm that since the flux is equal to the product of the particle concentration and the jump size and the latter is proportional to $\sqrt{D}$, then the stationary concentration ratio $\rho_R/\rho_L$ should be equal to $\sqrt{D_L/D_R}$, instead of the required $D_L/D_R$. The reason this argument is wrong is that instead of a perfectly sharp jump in the concentration, it actually changes from $\rho_L$ to $\rho_R$ over a length scale on the order of the length of the particle jump. Therefore the particle concentrations at the points where the jumps across the interface originate are not equal to the values far from the interface ($\rho_L$ and $\rho_R$, respectively). Note that, while this is true even when $r$ is continuously distributed, in the case $r=\pm 1$ there is an additional complication that affects the practical use of the BD algorithm: the particle concentration keeps varying wildly even far away from the interface, with the typical length scale of the oscillations again on the order of the particle jump size. As a consequence, the results depend strongly on how the data are binned. For this reason, using $r$ from a discrete set is best avoided in practical applications of the BD algorithm.

\section{Simulation Protocol}
\label{protocol}
The system is defined such that the left side extends from a wall at $x=-d/2$ to the interface at $x=0$ and the right side extends from the interface to a wall at $d/2$, where $d$ is the total length of the system. Thus the interface is placed equidistant between the walls. 
For all simulations described in this paper, the particle is initially placed at the interface. In the case of the MC II algorithm, when there is no site exactly at the interface, it starts at one of the sites immediately to the left and to the right of the interface with equal probabilities. 
Two types of simulations are performed. First, simulations with \textit{reflecting walls} at $x=-d/2$ and $x=d/2$ are conducted. To obtain the stationary particle distributions, relatively long simulations must be performed and thus we focus on a single case where $\eta_R / \eta_L =4$. Additional simulations with \textit{absorbing walls} are performed to investigate whether particles preferentially end up at one wall (herein referred to as an \emph{ultimate preferential direction}) and to measure the conditional mean first passage time (MFPT) to each wall. In most of these simulations, the viscosity on the left-hand side, $\eta_L$, is held fixed and the viscosity on the right-hand side is varied from $\eta_R=0.1\eta_L$ to $\eta_R=10\eta_L$; we also use $\eta_R/\eta_L=50$ or even, in Appendix A, $\eta_R/\eta_L=1000$.

For \textit{static} quantities, such as the stationary particle distributions and the ultimate preferential direction, only the ratio of the viscosities across the interface influences the results --- the actual values of the viscosities and the proportionality factor $\kappa$ in Eq.~(\ref{zetavseta}) do not matter. Indeed, for instance, looking at the expressions for the probabilities of the moves of the MC I algorithm [Eqs.~(\ref{ppmLR1}), (\ref{psLR1}), (\ref{ps1int}), (\ref{p0m1}), and (\ref{p0p1})], if $\eta_L$ and $\eta_R$ are both changed by some factor and at the same time the adjustable constant $\eta^{\I}_0$ is changed by the same factor, then the probabilities of the moves remain the same. What does change is the time scale, since changing $\eta^{\I}_0$ changes the time step $\tau_{\I}$, according to Eq.~(\ref{eta0I}), which corresponds to speeding up or slowing down the \textit{dynamics}. Since all approaches are expected to also give dynamically correct results such that the MFPTs and the decay of the transient behavior towards the stationary state should be consistent for any particular $\alpha$ value (with BD corresponding to $\alpha=0$ and LD to $\alpha=1$), a comparison of the dynamical quantities between the approaches is also meaningful. To make such comparisons, it is obviously necessary to know how the time steps of different algorithms are related.

It is convenient to choose the time step of the MC I algorithm as the unit of time and express all other times and time steps in terms of this unit. That is, numerically, the time step of MC I will be chosen to be
\begin{equation}
\tau_{\I}=1.\label{tauconv}
\end{equation}
Likewise, the mesh step of MC I is chosen as the unit of length ($a_{\I}=1$). In these units, we use the system size $d=50$, except for a single case in Appendix A. 
Thus there are 51 lattice sites at integer values of $x$ from $x=-25$ to $x=25$, and the interface at $x=0$ coincides with one of the sites as appropriate. The viscosity is expressed in units of $\eta^{\I}_0$; thus, numerically
\begin{equation}
\eta^{\I}_0=1.\label{etaconv}
\end{equation}
According to Eq.~(\ref{eta0I}), in these units
\begin{equation}
\frac{2kT}{\kappa}=1.\label{kappa}
\end{equation}

In MC II, we choose arbitrarily $a_{\II 0}=\sqrt{2}a_{\I}=\sqrt{2}$ in Eqs.~(\ref{al}) and (\ref{ar}). The lattice is placed in such a way that the interface is at distance $a_{\II}^L/2$ from the site adjacent to it on the left and at distance $a_{\II}^R/2$ from the site on the right. 
For the leftmost site, $-d/2\le x<-d/2+a_{\II}^L$, and for the rightmost site, $d/2-a_{\II}^R<x\le d/2$.  
It is convenient to express the viscosity in units of $\eta^{\II}_0$. However, for direct comparison with MC I the viscosity unit should be the same in both cases; thus we need $\eta^{\II}_0=\eta^{\I}_0$~($=1$). Using Eqs.~(\ref{eta0I}) and (\ref{eta0II}),
\begin{equation}
\frac{2kT\tau_{\II}}{\kappa a_{\II 0}^2(1-p_{\II s})}=\frac{2kT\tau_{\I}}{\kappa a_{\I}^2},
\end{equation}
so
\begin{equation}
\tau_{\II}=\tau_{\I}\frac{a_{\II 0}^2}{a_{\I}^2}(1-p_{\II s}),
\end{equation}
or, using the chosen values for $a_{\II 0}$ and $a_{\I}$ and keeping in mind that we have chosen the rest probability in MC II $p_{\II s}=2/3$ (as explained in Sec.~\ref{sec:MCII}),
\begin{equation}
\tau_{\II}=\frac{2}{3}\tau_{\I}=\frac{2}{3}.
\end{equation}

While choosing the values of the viscosities, we need to keep in mind that in MC I the viscosities cannot be smaller than $\eta^{\I}_0=1$. For the reflecting wall simulations, $\eta_L$ is set to 2, since, as we have mentioned, using a value too close to unity can introduce artifacts in the distribution function. Correspondingly, $\eta_R = 8$ for  $\eta_R/\eta_L =4$. For the absorbing wall case, for the simulations where we vary $\eta_R/\eta_L$ from 0.1 to 10, $\eta_L$ is set to 10 so that $\eta_R$ can be varied from 1 to 100. Using $\eta_R=1$ is not a problem in this case, since we are not interested in the details of the particle distribution or the FPT distribution, but only the MFPTs, in which case all oscillations and other artifacts average out. For $\eta_R/\eta_L=50$ and 1000, we use $\eta_L=2$, as in the reflecting case, so $\eta_R$ is respectively 100 and 2000.

For both MC algorithms, when the particle attempts to move left from the leftmost site or right from the rightmost site, in the case of absorbing boundaries it disappears and the simulation is stopped, while in the case of reflecting boundaries the move is rejected.

As for the MD simulations (BD and LD), we likewise use $d=50$ and treat the particle as pointlike in the sense that it can approach the walls infinitely closely. The random terms in the equations of motion are normally distributed. 
When attempting to cross a wall, the particle disappears in the absorbing case and is ``reflected'' in the reflecting case. 
The latter means that, for example, when the particle crosses the right wall at $d/2$ and would end up at $x=d/2+\Delta$, it is placed at $x=d/2-\Delta$ instead and in the LD case the sign of the particle velocity is changed; similarly for crossing the left wall. 
The equations of motion that are solved numerically [Eqs.~(\ref{SDE_D}) and (\ref{LD}), respectively] contain the diffusivity $D$. For comparison to MC, it needs to be expressed in terms of the viscosity given in the same units as in MC simulations, that is, in units of $\eta^{\I}_0$. Using Eqs.~(\ref{diffzeta}) and (\ref{kappa}), in these units
\begin{equation}
D=\frac{1}{2\eta}.\label{Dsimunits}
\end{equation}
For the BD time step $\delta t$, it is reasonable to choose a value similar to the time resolution of the MC schemes. We have chosen $\delta t=2$ for simulations with reflecting boundaries and $\delta t=20$ for simulations with absorbing boundaries. While the latter may appear too large, we have checked that using a ten times smaller step produces equivalent results. Ultimately, this is not surprising: What matters is that the time step is much smaller than the typical time that it takes the particle to reach the boundary starting at the interface, and this condition is satisfied. The situation with LD is somewhat more complicated. First of all, the initial velocity of the particle needs to be specified; it is chosen equal to either $+\sqrt{kT/m}$ or $-\sqrt{kT/m}$ at random. Second, there is one other relevant combination of parameters, $m/kT$, and the associated time scale, $mD/kT=m/2\eta kT$. It determines the inertia in the system, in particular, the correlation time of the particle velocity (or the time of relaxation to the terminal velocity). Since the intent is to model an overdamped system in which inertia is negligible, this time scale should be small, in any case smaller than the time it takes the particle to reach the wall from the interface. 
We chose $m/kT=4$ for reflecting boundaries and $m/kT=400$ for absorbing boundaries. In the least favorable case, $\eta=1$ for absorbing boundaries, the relaxation time is $t_r=m/2\eta kT=200$. 
The typical particle displacement over this time is $\sqrt{2Dt_r}=\sqrt{t_r/\eta}\approx 14$ which is less than a factor of 2 smaller than $d/2=25$.
While some discrepancies between LD and the MC and analytic results are indeed found at the lowest viscosities studied, good agreement is obtained for the majority of $\eta_R$ values.
The time step in LD should be much smaller than the relaxation time; we choose $\delta t=0.02$ for reflecting boundaries and $\delta t=0.2$ for absorbing boundaries. The least favorable case is the largest $\eta$ in our series for absorbing boundaries, $\eta=100$ (in the separate runs with $\eta_R$=2000 we do not use LD, only MC I). The relaxation time in this case is $t_r=2$, so this condition is satisfied. Note that the time step for LD is much smaller than for BD and in fact, LD is the most time-consuming algorithm of all those considered in this paper. This is expected, since in LD the time step should be much smaller than the relaxation time, which should itself be small; we note, however, that the efficiency of the LD algorithm could be improved without sacrificing its accuracy by using different $m/kT$ and $\delta t$ for different viscosity ratios.

\section{Results}

\subsection{Reflecting Walls: Distributions}
\label{refl}
In this section, we look at the particle concentration distributions for the case when the particles start at the interface. We first present the distributions for the Ito and isothermal calculi using the MC I and MC II algorithms, comparing them with the results obtained using BD (for Ito) and LD (for isothermal). We then compare the distributions for several different values of $\alpha$ using one of the MC algorithms (MC I). 
In all cases, we show the distributions at a time shortly after the start of the simulation and at a long time corresponding to the stationary state. 
In addition, for the Ito and isothermal calculi we also include one intermediate time, such that the distribution on the low-viscosity side has already reached the quasistationary state, but that on the high-viscosity side still resembles the short-time distribution.
For each curve, the total number of runs is 100000. 
In each run, a single particle is simulated and its position is determined at the times of interest. 
In MD simulations, the positions are binned with a unit bin size and the centers of the bins at half-integer $x$. 
The plotted values $\rho_i$ are proportional to the numbers of particles $n_i$ found at the given time at the site (or in the bin) $i$ and inversely proportional to the distance between the sites (width of the bin). 
The distributions are normalized so the area under the curve is unity (thus, in effect, these are probability density distributions). In particular, for MC I, BD and LD, since the distance between the sites (or the width of the bins) is unity, the normalization condition is
\begin{equation}
\sum\rho_i=1,
\end{equation}
where the sum runs over all sites (bins), and thus
\begin{equation}
\rho_i=\frac{n_i}{\sum_j n_j}.
\end{equation}
For MC II,
\begin{equation}
\sum_L \rho_i a_{\II}^L + \sum_R \rho_i a_{\II}^R = 1,
\end{equation}
where the first sum is over the sites in the left region and the second sum is over the sites in the right region, so
\begin{equation}
\rho_i=\left\{
\begin{array}{l l}
n_i/\left(a_{\II}^L\sum_j n_j \right)  & \ \text{in the left region,}\\
n_i/\left(a_{\II}^R\sum_j n_j \right) & \ \text{in the right region.}\\
\end{array}
\right.
\end{equation}

We also plot the ratio of the numbers of particles on the right and left sides as a function of time. When calculating this ratio, particles exactly at the interface (which is possible in the MC I case) are ignored.

\subsubsection{Ito calculus}
\label{ReflIto}
\begin{figure}[h!]
 	\centering
	\includegraphics[width=0.355\textwidth]{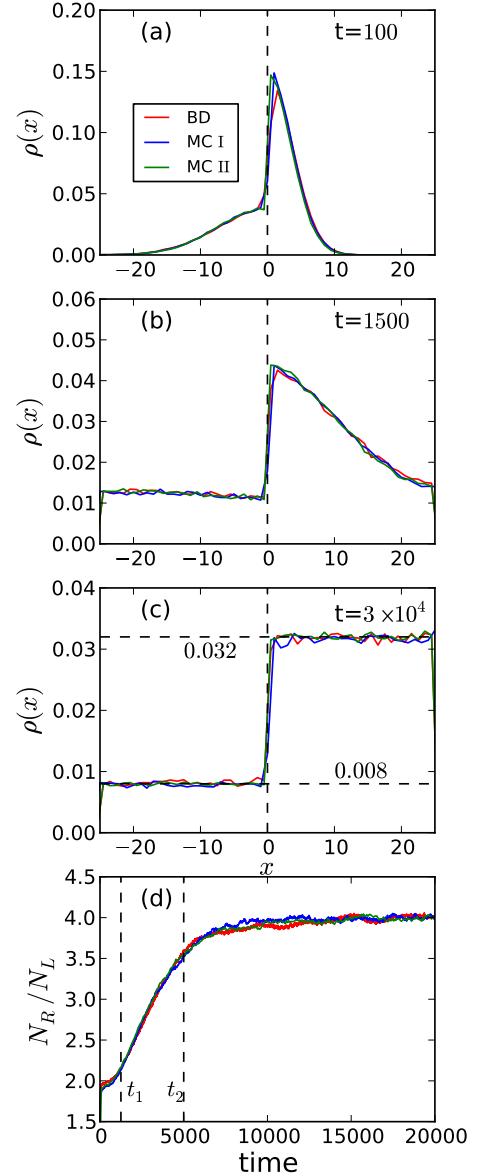}
	\caption{(color online) Results of reflecting wall simulations corresponding to the Ito calculus, for $\eta_L=2$, $\eta_R=8$ and $d=50$: particle concentration distributions, (a) at a short time, $t=100$, (b) at an intermediate time, $t=1500$, and (c) at a long time, $t=30000$,  as well as (d) the ratio of the number of particles on the right and left sides of the interface as a function of time, with the initial ($t_1$) and final ($t_2$) times of the transition period marked. All particles start at the interface in the middle of the system. In all plots, the results obtained using both MC methods, as well as BD simulations, are shown.}
	\label{fig:dist_Ito}
\end{figure}

The probability distributions $\rho$ as functions of position at three different times are shown in Fig. \ref{fig:dist_Ito}(a)--(c), for all algorithms yielding Ito calculus (MC I, MC II and BD).
In Fig. \ref{fig:dist_Ito}(a), the distribution shortly after starting the simulations, at time $t=100$, is shown. By this time, only very few particles reach the walls, so the distribution is essentially the same as it would be in infinite space, without the walls. In this case, theoretically the distribution is expected to follow
\begin{equation}
\rho(x,t)=\left\{
\begin{array}{l l}
\frac{1}{D_L}\frac{\sqrt{D_L D_R}} {\sqrt{D_L}+\sqrt{D_R}}\frac{1}{\sqrt{\pi t}}\exp\left(-\frac{x^2}{4D_L t}\right),\ & x<0,\\
\frac{1}{D_R}\frac{\sqrt{D_L D_R}} {\sqrt{D_L}+\sqrt{D_R}}\frac{1}{\sqrt{\pi t}}\exp\left(-\frac{x^2}{4D_R t}\right),\ & x>0.\\
\end{array}
\right.\label{ItoShortTime}
\end{equation}
This satisfies the diffusion equation~(\ref{diffpieces}) in both regions, as well as the jump condition (\ref{jump}) and the flux matching condition (\ref{fluxcont}), the latter because $\partial\rho/\partial x$ is zero both at $x=-0$ and $x=+0$.
Thus the parts of the distribution on the left side (L) and the right side (R) should both have a Gaussian form, but there should be a significant discontinuity in the distribution with a higher concentration of particles on the high viscosity side. These features are indeed reproduced in Fig.~\ref{fig:dist_Ito}(a). The discontinuity develops instantaneously (in a simulation, within a few steps from the beginning) and results from the trapping effect of high viscosity regions in the Ito formulation.
Figure \ref{fig:dist_Ito}(c) shows the distribution after a long time ($t=30000$) thus corresponding to the stationary state. While the transient piecewise-Gaussian shape has disappeared, the discontinuity persists yielding a step-function form. Recall the Ito stationary state condition [Eq.~(\ref{ItoStatio})] which, using $D \sim 1/\eta$, can be written as $\rho(x) / \eta(x) = {\rm const}$. Hence, as expected, the probability density on L, where the viscosity is low, is less than on R, where the viscosity is high, and the ratio is
\begin{equation}
\rho_R/\rho_L=\eta_R/\eta_L.
\end{equation}
Given the normalization condition $(\rho_L+\rho_R)d/2=1$,
\begin{eqnarray}
\rho_L&=&\frac{2}{d}\frac{\eta_L}{\eta_L+\eta_R},\\
\rho_R&=&\frac{2}{d}\frac{\eta_R}{\eta_L+\eta_R}.
\end{eqnarray}
For the case presented in Fig.~\ref{fig:dist_Ito} where $d=50$, $\eta_L=2$, $\eta_R=8$, we have $\rho_L = 0.008$, $\rho_R = 0.032$. 
These values are indicated on Fig. \ref{fig:dist_Ito}(c) where $\rho_R/\rho_L = 4$ as expected for $\eta_R/\eta_L = 4$. 
Note that at short times [as in Fig.~\ref{fig:dist_Ito}(a)], the ratio of the probability densities immediately to the right of the interface and immediately to the left of the interface is also the same. In fact, this is true at all times, as follows from Eq.~(\ref{jump}). The sharp drops in the distribution near the walls in Fig.~\ref{fig:dist_Ito}(c) are discretization and binning artifacts.

In Fig. \ref{fig:dist_Ito}(d), the ratio of the number of particles on R to L, denoted $N_R/N_L$, is plotted as a function of time. This ratio is equal to the ratio of the areas under the parts of the distribution curves on R and L.
Note that $N_R/N_L$ increases from a value around 2 and saturates at a value around 4.
Hence, there are always more particles on the high viscosity side: Under Ito conditions, the high viscosity side acts as a trap.
While the saturation value agrees with the stationary state condition of Ito calculus,
it is interesting that, beyond extremely short time behavior which depends more on the simulation algorithm than the physics,
the initial point for the increase during the transient portion is 2. Indeed, from Eq.~(\ref{ItoShortTime}) valid at short times, the ratio should be $\sqrt{\eta_R/\eta_L} = \sqrt{4} = 2$. This is the ratio of the heights of the half-Gaussians (which is equal to $D_L/D_R=\eta_R/\eta_L$) multiplied by the ratio of their widths (which is equal to $\sqrt{D_R/D_L}=\sqrt{\eta_L/\eta_R}$). At longer times, the particles begin to find the edge of the system and bounce back. Correspondingly, the system begins to move towards equilibrium where four times as many particles are found on the high viscosity side. The transition starts when the particles on the lower-viscosity side reach the boundary, which happens around time $t_1$ such that $2D_{\rm max}t_1=(d/2)^2$, thus $t_1=d^2/(8D_{\rm max})$, or, using Eq.~(\ref{Dsimunits}),
\begin{equation}
t_1=\frac{d^2\eta_{\rm min}}{4},\label{t1}
\end{equation}
where $D_{\rm max}$ is the higher of the two values of the diffusivity and $\eta_{\rm min}$ is the corresponding (lower of the two) value of the viscosity. The transition ends when the particles on the higher-viscosity site also reach the boundary, which occurs around
\begin{equation}
t_2=\frac{d^2\eta_{\rm max}}{4}.\label{t2}
\end{equation}
For our values of the parameters ($d=50$, $\eta_{\rm min}=\eta_L=2$, $\eta_{\rm max}=\eta_R=8$), $t_1=1250$ and $t_2$=5000.
These two times are marked on Fig.~\ref{fig:dist_Ito}(d). The time at which the distributions in Fig.~\ref{fig:dist_Ito}(a) are obtained is much smaller than $t_1$, thus corresponding to the situation before the transition starts, and the value of the ratio $N_R/N_L$ at that time is indeed close to two, according to Fig.~\ref{fig:dist_Ito}(d). On the other hand, the time at which the distributions in Fig.~\ref{fig:dist_Ito}(c) are obtained is much larger than $t_2$ and thus corresponds to the stationary situation after the transition has ended; the value of $N_R/N_L$ is close to four.

In Fig.~\ref{fig:dist_Ito}(b), we show the distribution at time $t=1500$, which is between $t_1$ and $t_2$. By time $t$ such that
\begin{equation}
t_1\ll t\ll t_2,\label{intermcond}
\end{equation}
particles on the low-viscosity side would have traversed that side many times, so the distribution on that side should be nearly flat resembling the stationary distribution, but on the other hand, very few particles would have reached the wall on the high-viscosity side, so the distribution on that side should still be close to the short-time Gaussian distribution. Even though in our case the ratio $t_2/t_1=\eta_R/\eta_L$ is only four, so, strictly speaking, satisfying Eq.~(\ref{intermcond}) is impossible, Fig.~\ref{fig:dist_Ito}(b) can still serve as a reasonable illustration of that regime. Note that the distribution on the low-viscosity side is not truly stationary, but rather \textit{quasi-stationary}, since there is a net outflow of particles to the high-viscosity side, as indicated by the gradual increase of $N_R/N_L$ between $t_1$ and $t_2$ [see Fig.~\ref{fig:dist_Ito}(d)]. This outflow is slow enough for the distribution to remain nearly flat, but the height of the distribution gradually decreases. In essence, the height of the plateau on the low-viscosity side adiabatically follows the height of the peak on the high-viscosity side (which decreases as this peak broadens), so that the boundary condition (\ref{jump}) is satisfied at all times. On the other hand, this outflow is large enough that the slope of the distribution function is significantly nonzero immediately to the right of the interface in Fig.~\ref{fig:dist_Ito}(c). However, the viscosity ratio is rather small here and also $t$ is close enough to $t_2$ that particles have already started to reach the right wall (note a nonzero particle concentration next to it). As we argue in Appendix A, for a very high viscosity ratio and when the condition (\ref{intermcond}) is truly satisfied, the derivative at $x=+0$ is negligible and the distribution on the high-viscosity side is a Gaussian.

Taking into account the general character of the distribution at intermediate times between $t_1$ and $t_2$, we can also obtain approximately the time dependence of the ratio $N_R/N_L$ in this time interval. 
The number of particles on R, $N_R$, is roughly proportional to $\rho(+0,t)$ times the width of the distribution on that side, which is $\propto\sqrt{t}$. 
The number of particles on L, $N_L$, is proportional to $\rho(-0,t)$ times the length of that part of the system, $d/2$, which is time-independent. Thus $N_R/N_L\propto[\rho(+0,t)/\rho(-0,t)]\sqrt{t}$, and, given that, according to Eq.~(\ref{jump}), the ratio in the square brackets is time-independent,
\begin{equation}
\frac{N_R}{N_L}\propto\sqrt{t}.\label{NRoverNL}
\end{equation}

This agreement between the theory at $\alpha=0$ and the Brownian dynamics results confirms that the BD simulations correspond to Ito calculus.
Further, note the overall good agreement between the BD simulations and both MC approaches for all the results shown in Fig. \ref{fig:dist_Ito}.
Considering factors such as discretization, binning, and noise, the fact that the differences are small between the data sets is a convincing validation of the two MC approaches we have developed to simulate Ito calculus.

\subsubsection{Isothermal calculus}

\begin{figure}[h!]
 	\centering
	\includegraphics[height=0.77\textheight]{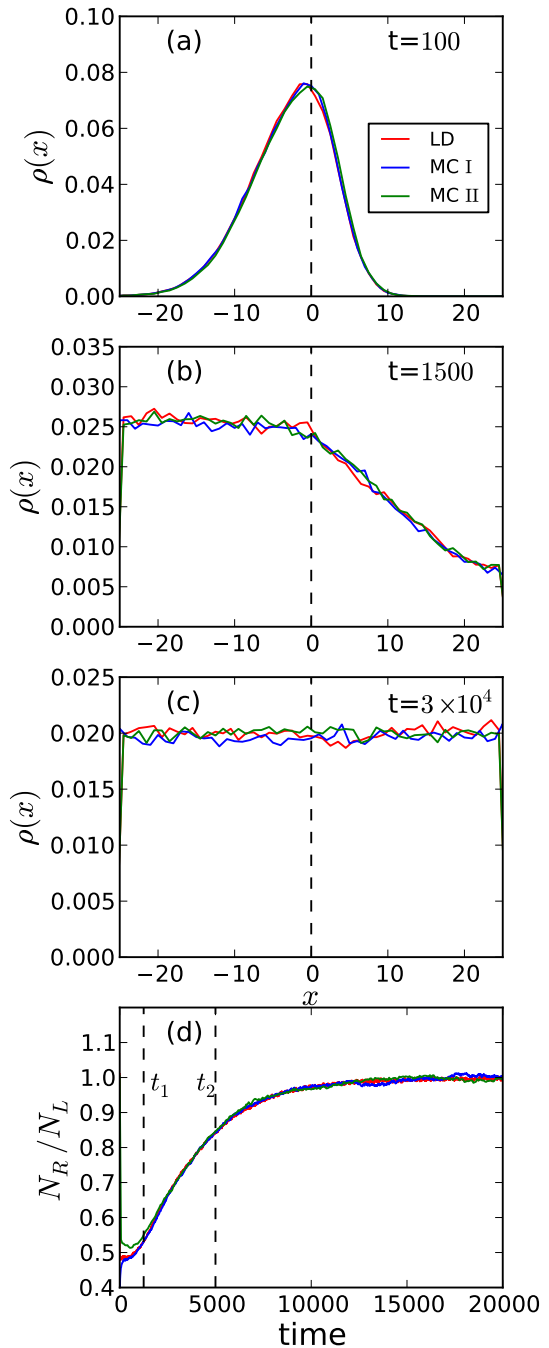}
	\caption{(color online) Results of reflecting wall simulations corresponding to the isothermal calculus, for $\eta_L=2$, $\eta_R=8$ and $d=50$: particle concentration distributions, (a) at a short time, $t=100$, (b) at an intermediate time, $t=1500$, and (c) at a long time, $t=30000$, as well as (d) the ratio of the number of particles on the right and left sides of the interface as a function of time, with the initial ($t_1$) and final ($t_2$) times of the transition period marked. All particles start at the interface in the middle of the system. In all plots, the results obtained using both MC methods, as well as LD simulations, are shown.}
	\label{fig:dist_iso}
\end{figure}

The equivalent plots for all algorithms yielding the isothermal calculus (MC I, MC II and LD) are shown in Fig. \ref{fig:dist_iso}. Note that the distributions are now continuous across the interface, in agreement with Eq.~(\ref{jump}) for $\alpha=1$.
For the short time behavior shown in Fig. \ref{fig:dist_iso}(a), where, again, $t=100$, the Gaussian-like distribution on the left side is wider than that on the right side, as in the Ito case. Theoretically, the distribution at short times is given by
\begin{equation}
\rho(x,t)=\left\{
\begin{array}{l l}
\frac{1}{\sqrt{D_L}+\sqrt{D_R}}\frac{1}{\sqrt{\pi t}}\exp\left(-\frac{x^2}{4D_L t}\right),\ & x<0,\\
\frac{1}{\sqrt{D_L}+\sqrt{D_R}}\frac{1}{\sqrt{\pi t}}\exp\left(-\frac{x^2}{4D_R t}\right),\ & x>0,\\
\end{array}
\right.\label{IsoShortTime}
\end{equation}
which is the same as Eq.~(\ref{ItoShortTime}) for the Ito case, except for the prefactors that ensure the continuity across the interface.
At long times, in the stationary state, there is an equal probability to find a particle anywhere in the system, as is seen in Fig. \ref{fig:dist_iso}(c) ($t=30000$). This is consistent with the isothermal stationary state condition [Eq.~(\ref{IsoStatio})]. The stationary value of the density is obviously given by
\begin{equation}
\rho=\frac{1}{d}=0.02.
\end{equation}
The time dependence of the ratio of the number of particles on R over L is shown in Fig.~\ref{fig:dist_iso}(d).
Here, the transient behavior begins from the ratio of approximately 0.5. That is, given the bias at the interface to end up on the lower viscosity side, very rapidly there is a greater number of particles on L, and, contrary to the Ito results where there are twice as many particles on the high viscosity side near the beginning, for the isothermal results there are half as many particles on the high viscosity side. This ratio can again be obtained by considering the peak height and width. Here, the peak height is the same on both sides while the ratio of the widths is again $\sqrt{\eta_L/\eta_R}$.
Correspondingly, the ratio is $N_R/N_L = \sqrt{\eta_L}/\sqrt{\eta_R} = 1/2$.
As the system moves towards equilibrium, $N_R/N_L$ again climbs and plateaus,
but now it saturates at 1.0 in accord with the isothermal stationary state condition. The transition between the initial and the final values again starts around $t_1$ given by Eq.~(\ref{t1}) and ends around $t_2$ given by Eq.~(\ref{t2}). At times between $t_1$ and $t_2$, the distribution is nearly flat (quasi-stationary) on the low-viscosity side and Gaussian-like on the high-viscosity side, again, as in the Ito case [see Fig.~\ref{fig:dist_iso}(b) for $t=1500$]. However, unlike in the Ito case, we show in Appendix A that even for a large viscosity ratio and $t_1\ll t\ll t_2$, the slope at $+0$ is significantly nonzero, and, in fact, the interface lies at the inflection point of the Gaussian-like curve rather than at its maximum. Despite this, the crude arguments leading to Eq.~(\ref{NRoverNL}) still apply, so we expect that time dependence to still be approximately valid here. We compare the time dependences of $N_R/N_L$ for different calculi in the next section.

In all four plots, again, there is good agreement between all three algorithms.

\subsubsection{Comparison between calculi}
\begin{figure}[h!]
 	\centering
	\includegraphics[height=0.745\textheight]{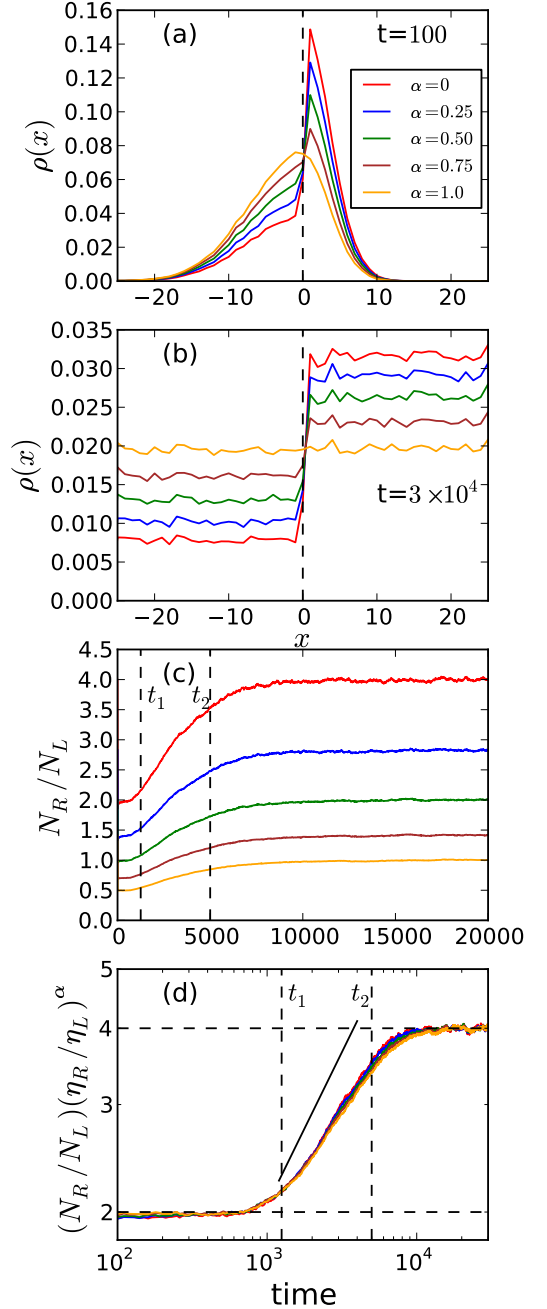}
	\caption{(color online) Results of reflecting wall simulations using the MC I algorithm for different calculi, with the values of $\alpha$ specified in the legend, for $\eta_L=2$, $\eta_R=8$ and $d=50$: particle concentration distributions, (a) at a short time, $t=100$, and (b) at a long time, $t=30000$; (c) and (d) the ratio of the number of particles on the R and L sides of the interface [rescaled in (d) as indicated] as a function of time, with the initial ($t_1$) and final ($t_2$) times of the transition period marked. The rescaling factor in (d) is chosen so that the values coincide at both small and large times, yet in the intermediate region small differences are observed. The straight line indicates the predicted $\sim t^{1/2}$ behavior. All particles start at the interface in the middle of the system.}
	\label{fig:dist_all}
\end{figure}

In Fig.~\ref{fig:dist_all}, we compare the results for the particle distributions and the ratio $N_R/N_L$ for several different calculi (values of $\alpha$), including the already considered Ito ($\alpha=0$) and isothermal ($\alpha=1$) limit cases, using the MC I algorithm. Essentially identical results are obtained using MC II (not shown). 
Figure~\ref{fig:dist_all}(a) shows the distributions at $t=100$ and Fig.~\ref{fig:dist_all}(b) presents the distributions at $t=30000$. 
At short times [Fig.~\ref{fig:dist_all}(a)], the distribution is always piecewise-Gaussian with
\begin{equation}
\rho(x,t)=\left\{
\begin{array}{l l}
\frac{D_L^{\alpha-1}} {D_L^{\alpha-1/2}+D_R^{\alpha-1/2}}\frac{1}{\sqrt{\pi t}}\exp\left(-\frac{x^2}{4D_L t}\right),\ & x<0,\\
\frac{D_R^{\alpha-1}} {D_L^{\alpha-1/2}+D_R^{\alpha-1/2}}\frac{1}{\sqrt{\pi t}}\exp\left(-\frac{x^2}{4D_R t}\right),\ & x>0.\\
\end{array}
\right.\label{AllShortTime}
\end{equation}
There is a jump at the interface in all cases, except $\alpha=1$, in agreement with Eq.~(\ref{jump}). At long times [Fig.~\ref{fig:dist_all}(b)], the distribution is piecewise-uniform, with
\begin{eqnarray}
\rho_L&=&\frac{2}{d} \times \frac{\eta_L^{1-\alpha}}{\eta_L^{1-\alpha}+\eta_R^{1-\alpha}},\\
\rho_R&=&\frac{2}{d}\times \frac{\eta_R^{1-\alpha}}{\eta_L^{1-\alpha}+\eta_R^{1-\alpha}},
\end{eqnarray}
which follows from Eq.~(\ref{equil}) for the stationary distribution. Again, there is a jump for $\alpha\ne 1$. In Fig.~\ref{fig:dist_all}(c), the ratio $N_R/N_L$ is plotted. The short-time value of this ratio is
\begin{equation}
\frac{N_R}{N_L}=\left(\frac{\eta_R}{\eta_L}\right)^{1/2-\alpha},\label{shorttime}
\end{equation}
which is the product of the ratio of the peak widths [$(\eta_R/\eta_L)^{-1/2}$) and the ratio of the peak heights [$(\eta_R/\eta_L)^{1-\alpha}$]. The long-time (stationary-state) value of $N_R/N_L$ is
\begin{equation}
\frac{N_R}{N_L}=\left(\frac{\eta_R}{\eta_L}\right)^{1-\alpha}.\label{longtime}
\end{equation}
For $0\le\alpha<1/2$, $N_R/N_L$ is always larger than 1, so there are always more particles on the higher-viscosity R side than on L. For $\alpha=1/2$, the ratio is one initially, but then grows above unity. For $\alpha=1$, there are always more particles on the \textit{lower-viscosity} site, but the ratio approaches unity from below as $t\to\infty$. Finally, for $1/2<\alpha<1$, there are more particles on the low-viscosity side initially, but the ratio crosses unity at a finite time and eventually there are more particles on the higher-viscosity side. The transition between the short-time and the long-time values of $N_R/N_L$ is roughly between $t_1$ and $t_2$ given by Eqs.~(\ref{t1}) and (\ref{t2}), and the time dependence should still be roughly given by Eq.~(\ref{NRoverNL}).

Note that at both small and large times, if the particle distribution on L is multiplied by $(\eta_R/\eta_L)^{1-\alpha}$, then the resulting function is the same for all $\alpha$: This rescaling eliminates the jump at zero, and the shapes of the distributions in a given region and at any given time are calculus-independent, both at short times (when they are Gaussian with the calculus-independent width) and at long times (when they are flat). 
A reasonable question then is if this is also true at intermediate times and thus at all times. 
The answer, perhaps somewhat surprisingly, is \textit{no}. 
This can be seen by comparing the shapes of the intermediate-time distributions for the Ito [Fig.~\ref{fig:dist_Ito}(b)] and isothermal [Fig.~\ref{fig:dist_iso}(b)] cases that are indeed somewhat different, and, as mentioned, the difference is significant in the high-viscosity-ratio limit (see Appendix A). 
This difference is also easy to explain: 
If we take, for example, the solution for the Ito case and multiply its left half by $\eta_R/\eta_L$, the resulting function will satisfy the diffusion equation (\ref{diffpieces}) and the jump condition (\ref{jump}) for $\alpha=1$, but \textit{not} the flux continuity condition (\ref{fluxcont}), since only the left-hand side of the latter will be multiplied by $\eta_R/\eta_L$, but not the right-hand side. 
The cases $t\to 0$ and $t\to\infty$ are exceptions, because in these limits $\partial\rho/\partial x$ is zero at both $x=-0$ and $x=+0$. 
However, for intermediate times this cannot be the case, since there \textit{must} be some net flux between the two sides, or otherwise $N_R/N_L$ would remain constant. 
As a consequence, while the curves in Fig.~\ref{fig:dist_all}(c) for different $\alpha$ do look similar, and, in particular, the transition between the low-time and the high-time values of $N_R/N_L$ occurs over \textit{roughly} the same time interval and according to \textit{approximately} the same $\sim t^{1/2}$ dependence in all cases, these curves are not perfect rescaled copies of each other: Multiplying these functions by $(\eta_R/\eta_L)^{\alpha}$ so they match at $t\to 0$ and $t\to\infty$ does not make them coincide everywhere, as illustrated in Fig.~\ref{fig:dist_all}(d), where small differences between the rescaled curves around $t_2$ are apparent. 
That plot also compares the curves with the $t^{1/2}$ dependence predicted for large viscosity ratios; the agreement is reasonable given that in our case the ratio $\eta_R/\eta_L=4$ is not large.

\subsection{Absorbing Walls: Preferential Direction}

We now describe the results of simulations with \textit{absorbing} walls, that is, each single-particle simulation run is stopped once the particle reaches one of the two walls. 
For the Ito and isothermal calculi, we first look at whether the particles prefer to end up on one of the walls (i.e., whether there is an \textit{ultimate preferential direction}) and find the conditional MFPTs, separately for particles reaching each wall. 
The \textit{ultimate} preference is then compared to the preference of the particle to spend more time on one or the other side, which is analyzed in two different ways. 
Unlike the case of reflecting boundaries, where a single set of viscosities was considered, here we vary the ratio of the viscosities, with the viscosity on L kept constant.
For each set of viscosities, we carry out 10000 single-particle runs, again starting at the interface, as described in Sec.~\ref{protocol}. 
For the Ito calculus we also obtain the full conditional FPT distributions for a single representative viscosity ratio using $10^6$ runs.

\subsubsection{Ito calculus}
\label{absIto}

\begin{figure}[h!]
 	\centering
	\includegraphics[width=0.4\textwidth]{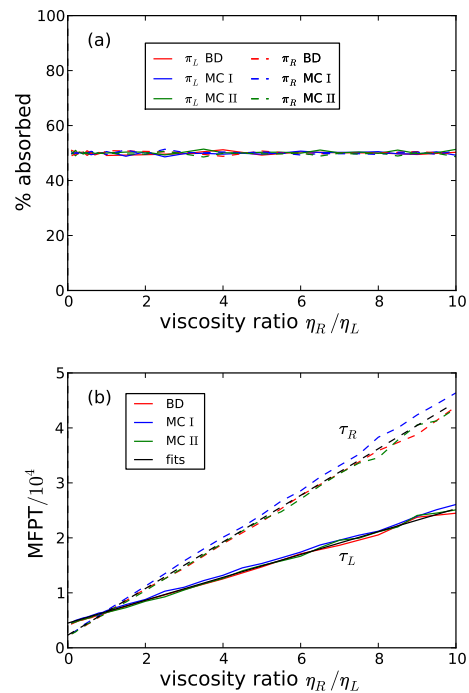}
	\caption{(color online) Results of absorbing wall simulations corresponding to the Ito calculus, for $\eta_L=10$, $d=50$, and varying $\eta_R$: (a) the percentage of particles absorbed by L (solid lines) and R (dashed lines) walls; (b) the MFPT under the condition that the particle is absorbed by the left wall (solid lines) and by the right wall (dashed lines). All particles start at the interface in the middle of the system. In both plots, the results obtained using both MC methods, as well as BD simulations, are shown. Black lines in (b) are linear fits to the averages of all three methods.}
	\label{fig:pd_ito}
\end{figure}

Results for the absorbing wall simulations in the case of Ito calculus are shown in Fig. \ref{fig:pd_ito}. 
Figure \ref{fig:pd_ito}(a) displays the fraction of events absorbed on R, $\pi_R$ (dashed lines) and L, $\pi_L$ (solid lines) for all three algorithms. 
For each algorithm at any viscosity ratio, the number absorbed at L equals the number absorbed at R within error. Thus, under Ito conditions, there is no ultimate preferential direction in the system. 
This result may seem paradoxical at first, given that for large $\eta_R/\eta_L$ there are much fewer particles on L than on R \textit{at any given moment of time}, 
which we have seen for reflecting boundaries [Fig. \ref{fig:dist_Ito}(d)] and which turns out to be true for absorbing boundaries as well. 
The primary reason for the equality in Fig. \ref{fig:pd_ito} (a) is that particles diffuse faster on L, so even a smaller concentration creates a significant flux towards the left wall.

\begin{figure}[h!]
 	\centering
	\includegraphics[width=0.4\textwidth]{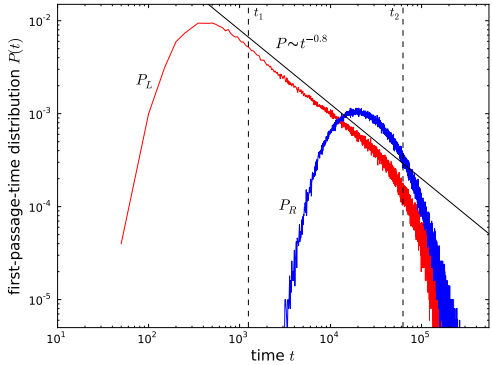}
	\caption{(color online) The conditional FPT distributions for the left and right walls for $\eta_L=2$, $\eta_R=100$, $d=50$ obtained using the MC I algorithm. Times $t_1$ and $t_2$ given by Eqs.~(\ref{t1}) and (\ref{t2}), respectively, are marked.}
	\label{fig:distLR}
\end{figure}

The conditional MFPTs to each wall (which we denote $\tau_L$ and $\tau_R$) are plotted in Fig. \ref{fig:pd_ito}(b). 
As would be expected, the time to the lower viscosity side is always less than the time to the high viscosity side with an intersection corresponding to equal times at $\eta_R / \eta_L = 1$. A less obvious fact is that the two times are always of the same order of magnitude even when the viscosities are very different.
In fact, both dependences are linear as demonstrated by straight-line fits (black lines) to the averages of all three algorithms, and the slope for the R data is, within the uncertainty, two ($2.04\pm 0.07$) times that of the slope for the L data, so the ratio of the two times approaches two as $\eta_R/\eta_L\to\infty$. 
This may seem surprising, given that the time $t_1$ [Eq.~(\ref{t1})] around which a significant (not exponentially small) number of particles first reaches the left wall is much smaller in this limit than the corresponding time $t_2$ [Eq.~(\ref{t2})] for the right wall. 
The explanation is that, while it is indeed true that the particle flow to the wall on the low-viscosity side starts at time $\sim t_1$, it continues until time $\sim t_2$, as the particles keep ``leaking'' from the high-viscosity side until those that have not been absorbed by the wall on the low-viscosity side are absorbed by the other wall. Even if the means are of the same order of magnitude, the conditional first-passage time \textit{distributions} $P_{L,R}(t)$ are very different when the viscosities are very different. For the wall on the low-viscosity side, the flux reaches its maximum very fast, by time $\sim t_1$ and then decreases very gradually (as a power law) until time $\sim t_2$, after which it decays exponentially. 
But for the wall on the high-viscosity side, the flux is essentially zero until time $\sim t_2$ and then rises and decays relatively quickly. 
This behavior of the distributions is illustrated in Fig.~\ref{fig:distLR} obtained for a large viscosity ratio ($\eta_L=2$, $\eta_R=100$, so $\eta_R/\eta_L=50$) using the MC I method with the waiting time selected from a distribution as given by Eq.~(\ref{ndistr}).
The best fit to the power-law region has the exponent $\approx -0.8$; we show in Appendix A that in the limit of a very large viscosity ratio this exponent is $-0.5$.

To derive the properties revealed in Fig.~\ref{fig:pd_ito}, consider the variant of the MC I algorithm with all steps successful (no staying put) and the time increment at each step $t_w$ given by Eq.~(\ref{allsucc}) (see Sec.~\ref{MCIallsucc}). Recall that in the Ito formulation, there is no bias at the interface [from Eqs.~(\ref{pminussucc}) and (\ref{pplussucc}), $p'^0_{\I -}=p'^0_{\I +}=1/2$], or anywhere else in the system. Hence, the trajectory of a particle is just an unbiased random walk, except for the fact that the time increment at each step changes depending on what side of the interface the particle is at. This exception, of course, does not change the fact that this random walk is equally likely to first reach the left or the right wall, since it starts at equal distances from them, which explains the result in Fig.~\ref{fig:pd_ito}(a). To get insight into the results for the MFPTs in Fig.~\ref{fig:pd_ito}(b), we use the fact that if a particle is released equidistant between two absorbing walls that are at distance $d$ from each other and this particle does a random walk with step size unity eventually reaching the right-hand wall, then on average it spends
\begin{equation}
n_{\rm same}=\frac{d^2}{6}
\end{equation}
steps on the right-hand side and
\begin{equation}
n_{\rm opp}=\frac{d^2}{12}
\end{equation}
steps on the left-hand side (see Appendix B). Then
\begin{equation}
\tau_R=t_{\I w}^L n_{\rm opp}+t_{\I w}^R n_{\rm same}=\frac{d^2 \tau_{\I}}{12}\frac{\eta_L}{\eta^{\I}_0}[1+2(\eta_R/\eta_L)],
\end{equation}
and similarly,
\begin{equation}
\tau_L=t_{\I w}^L n_{\rm same}+t_{\I w}^R n_{\rm opp}=\frac{d^2 \tau_{\I}}{12}\frac{\eta_L}{\eta^{\I}_0}[2+(\eta_R/\eta_L)],
\end{equation}
where $t_{\I w}^L$ and $t_{\I w}^R$ are the time steps [given by Eq.~(\ref{allsucc})] on L and R, respectively, or, using the conventions (\ref{tauconv}) and (\ref{etaconv}),
\begin{eqnarray}
\tau_R&=&\frac{d^2\eta_L}{12}[1+2(\eta_R/\eta_L)],\label{tauR}\\
\tau_L&=&\frac{d^2\eta_L}{12}[2+(\eta_R/\eta_L)].\label{tauL}
\end{eqnarray}
For $\eta_L={\rm const}$, the dependence of both times on $\eta_R/\eta_L$ is indeed linear, and the ratio of the slopes is two, in agreement with the data in Fig.~\ref{fig:pd_ito}(b). In our case ($d=50$, $\eta_L=10$),
\begin{eqnarray}
\tau_R&\approx&2083+4167(\eta_R/\eta_L),\\
\tau_L&\approx&4167+2083(\eta_R/\eta_L).
\end{eqnarray}
This should be compared with the fits to the data (black lines in Fig.~\ref{fig:pd_ito}), which are
\begin{eqnarray}
\tau_R^{\rm fit}&=&(2.35\pm 0.03)\times 10^3\nonumber\\
& &+(4.26\pm 0.09)\times 10^3(\eta_R/\eta_L),\\
\tau_L^{\rm fit}&=&(4.58\pm 0.04)\times 10^3\nonumber\\
& &+(2.09\pm 0.06)\times 10^3(\eta_R/\eta_L).
\end{eqnarray}
The slopes are indeed in agreement; the small disagreement in the intercepts (about 10\%) is likely due to discretization effects.
In the limit $\eta_R/\eta_L\to\infty$, Eqs.~(\ref{tauR}) and (\ref{tauL}) become
\begin{eqnarray}
\tau_R&\approx&\frac{d^2\eta_R}{6}=\frac{2}{3}t_2,\\
\tau_L&\approx&\frac{d^2\eta_R}{12}=\frac{1}{3}t_2,
\end{eqnarray}
where $t_2$ is given by Eq.~(\ref{t2}).
This shows that in this limit both times become independent of the lower viscosity and in particular, both are $\sim t_2$.
We discuss the relation between this result and the FPT distribution in Appendix A.

\subsubsection{Isothermal calculus}
\begin{figure}[h!]
 	\centering
	\includegraphics[width=0.4\textwidth]{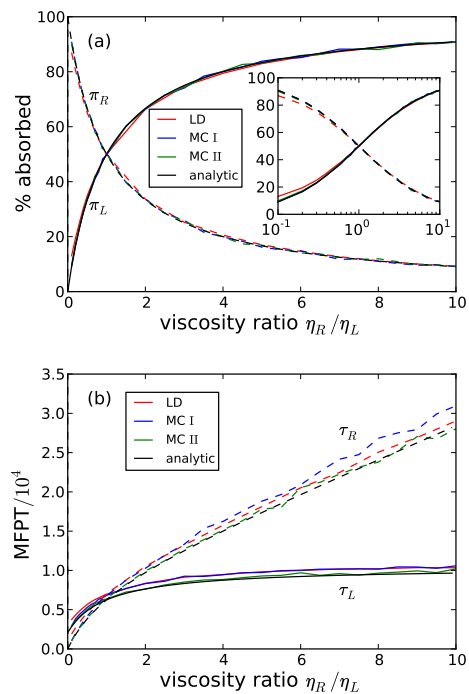}
	\caption{(color online) Results of absorbing wall simulations corresponding to the isothermal calculus, for $\eta_L=10$, $d=50$, and varying $\eta_R$: (a) the fraction of particles $\pi_{L,R}$ absorbed by L (solid lines) and R (dashed lines) walls; (b) the MFPT under the condition that the particle is absorbed by the left wall (solid lines) and by the right wall (dashed lines). Black lines in both plots are theoretical dependences derived in the text. All particles start at the interface in the middle of the system. In both plots, the results obtained using both MC methods, as well as LD simulations, are shown.}
	\label{fig:pd_iso}
\end{figure}

The same data for the case of isothermal calculus are shown in Fig. \ref{fig:pd_iso}. 
There are significant quantitative and qualitative differences compared to the case of Ito calculus. First, examining Fig. \ref{fig:pd_iso}(a), there is a clear ultimate preferential direction in the system.
When $\eta_R / \eta_L = 0.1$, approximately 90\% of the particles are absorbed at the R wall.
Conversely, when $\eta_R / \eta_L = 10$, 90\% of the particles are absorbed at the L wall.
Plotting the data with a log scale for the viscosity ratio [inset to Fig. \ref{fig:pd_iso}(a)] demonstrates that this symmetry holds for all simulated viscosity ratios:
\begin{equation}
\pi_R(\eta_R/\eta_L) = \pi_L(\eta_L/\eta_R),
\label{symm}
\end{equation}
where $\pi_{L,R}(z)$ are the probabilities of absorption at the left (right) wall when the ratio of viscosities on R and on L is $z$.
This symmetry is a necessary outcome considering that in the algorithms, it is only the ratio of viscosities, and not the absolute value, which influences the dynamics
(beyond simply making all events take longer).
The ultimate preferential direction under isothermal conditions arises from the bias at the interface favoring the low viscosity side.
Since this bias increases with a larger disparity in viscosities, the strength of the preference in direction also grows.
It must, however, saturate at 100\% resulting in the behavior shown in Fig. \ref{fig:pd_iso}(a).

The MFPTs, shown in Fig. \ref{fig:pd_iso}(b), are also quite different from those corresponding to the Ito condition.
Again, as expected, the time to the lower viscosity side is lower than the time to the high viscosity side with an intersection at $\eta_R / \eta_L = 1$.
However, the dependence across the entire viscosity range is clearly not linear.
In fact, considering data for which  $\eta_R / \eta_L > 1$, while $\tau_R$ does increase linearly with increasing $\eta_R / \eta_L$ (albeit only when this ratio is large), $\tau_L$ appears to saturate.

To understand the results in Fig.~\ref{fig:pd_iso}, we turn again to the variant of the MC I algorithm with all steps successful. 
For generality, we consider an arbitrary $\alpha$ and then specialize to $\alpha=1$. 
For $\alpha\ne 0$, the segments of the particle trajectory between visits to the interface site are still random walks, but when the particle is at the interface, it chooses to continue on L or on R with the probabilities given by Eqs.~(\ref{pminussucc}) and (\ref{pplussucc}), respectively. Each such trajectory segment is contained entirely either within the left half or within the right half and for brevity, we refer to these groups of segments as left and right segments. Since the probabilities to reach the respective wall during the left and right segments are equal, the total probability to reach the left wall, $\pi_L$, is equal to the fraction of left segments, which in its turn is equal to the probability to start a left segment while at the interface, which is $p'^0_{\I -}$. Correspondingly, the probability to reach the right wall, $\pi_R$, is $p'^0_{\I +}$. So
\begin{eqnarray}
\pi_L&=&\frac{(\eta_R/\eta_L)^{\alpha}}{1+(\eta_R/\eta_L)^{\alpha}},\label{piL}\\
\pi_R&=&\frac{1}{1+(\eta_R/\eta_L)^{\alpha}}.\label{piR}
\end{eqnarray}
Note that $\pi_L\to 1$ for $\eta_R/\eta_L\to\infty$ whenever $\alpha>0$, whereas for $\alpha=0$, $\pi_L\equiv 1/2$, as we have seen in the previous section. In the isothermal case ($\alpha=1$)
\begin{eqnarray}
\pi_L&=&\frac{\eta_R/\eta_L}{1+\eta_R/\eta_L},\label{piLIso}\\
\pi_R&=&\frac{1}{1+\eta_R/\eta_L}.\label{piRIso}
\end{eqnarray}
These functions are plotted in Fig.~\ref{fig:pd_iso}(a), and agreement with the simulation data is excellent. 
Equations~(\ref{piLIso}) and (\ref{piRIso}) satisfy the symmetry condition (\ref{symm}). At $\eta_R/\eta_L=0.1$, $\pi_R=10/11\approx 91\%$, while at $\eta_R/\eta_L=10$, $\pi_L=10/11\approx 91\%$, again, in agreement with the simulations.

To calculate the MFPTs, we use the result obtained in Appendix B that for a random walk that is unbiased, except at the interface (where it chooses to go left with probability $p'^0_{\I -}$ or right with probability $p'^0_{\I +}$), the mean number of steps that the walk absorbed by the right wall spends on R is
\begin{equation}
n_{RR}=\frac{d^2}{12}(1+2p'^0_{\I +}),
\end{equation}
while on L it spends on average
\begin{equation}
n_{RL}=\frac{d^2}{6}p'^0_{\I -}
\end{equation}
steps. Then the MFPT for the walk ending on the right wall is
\begin{eqnarray}
\tau_R&=&t_{\I w}^L n_{RL}+t_{\I w}^R n_{RR}\label{tauRgen}\\
&=&\frac{d^2 \eta_L}{12}\times\frac{(\eta_R/\eta_L)^{1+\alpha}+3(\eta_R/\eta_L)+2(\eta_R/\eta_L)^{\alpha}}{1+(\eta_R/\eta_L)^{\alpha}},\nonumber
\end{eqnarray}
where we have used Eq.~(\ref{allsucc}) for $t_{\I w}^L$ and $t_{\I w}^R$ and conventions (\ref{tauconv}) and (\ref{etaconv}). For walks ending on the left wall, analogously there are on average
\begin{equation}
n_{LL}=\frac{d^2}{12}(1+2p'^0_{\I -})
\end{equation}
steps on L and
\begin{equation}
n_{LR}=\frac{d^2}{6}p'^0_{\I +}
\end{equation}
steps on R, and then
\begin{eqnarray}
\tau_L&=&t_{\I w}^L n_{LL}+t_{\I w}^R n_{LR}\label{tauLgen}\\
&=&\frac{d^2 \eta_L}{12}\times\frac{2(\eta_R/\eta_L)+3(\eta_R/\eta_L)^{\alpha}+1}{1+(\eta_R/\eta_L)^{\alpha}}.\nonumber
\end{eqnarray}
In particular, in the isothermal case ($\alpha=1$),
\begin{eqnarray}
\tau_R&=&\frac{d^2 \eta_L}{12}\times\frac{(\eta_R/\eta_L)^2+5(\eta_R/\eta_L)}{1+(\eta_R/\eta_L)},\label{tauRIso}\\
\tau_L&=&\frac{d^2 \eta_L}{12}\times\frac{5(\eta_R/\eta_L)+1}{1+(\eta_R/\eta_L)}.\label{tauLIso}
\end{eqnarray}
Equations~(\ref{tauRIso}) and (\ref{tauLIso}) with $\eta_L=10$ and $d=50$ are plotted in Fig.~\ref{fig:pd_iso}(b). The agreement with the simulations is very good; discrepancies can be attributed to discretization effects as well as inertial effects in the LD case.
Equations~(\ref{tauR}) and (\ref{tauL}) are recovered from Eqs.~(\ref{tauRgen}) and (\ref{tauLgen}) for $\alpha=0$. When $\eta_R/\eta_L$ is large, Eqs.~(\ref{tauRgen}) and (\ref{tauLgen}) give
\begin{equation}
\tau_R\approx\left\{
\begin{array}{l l}
\frac{d^2\eta_R}{6}=\frac{2t_2}{3} & {\rm \ for\ }\alpha=0,\\
\frac{d^2\eta_R}{12}=\frac{t_2}{3} & {\rm \ for\ }0<\alpha\le 1,\\
\end{array}
\right.
\end{equation}
\begin{equation}
\tau_L\approx\left\{
\begin{array}{l l}
\frac{d^2\eta_R}{12}=\frac{t_2}{3} & {\rm \ for\ }\alpha=0,\\
\frac{d^2\eta_L^{\alpha}\eta_R^{1-\alpha}}{6}=\frac{2t_1^{\alpha}t_2^{1-\alpha}}{3} & {\rm \ for\ }0<\alpha<1,\\
\frac{5d^2\eta_L}{12}=\frac{5t_1}{3} & {\rm \ for\ }\alpha=1.\\
\end{array}
\right.
\label{tauLgenlim}
\end{equation}
For $\eta_L={\rm const}$, this always gives a linear dependence on $(\eta_R/\eta_L)$ for $\tau_R$, regardless of $\alpha$, and, unsurprisingly, $\tau_R\sim t_2$, again for any $\alpha$. 
For $\tau_L$, the dependence is $\propto (\eta_R/\eta_L)^{1-\alpha}$, which is linear for $\alpha=0$, reproducing the result of the previous section, sublinear for $0<\alpha<1$, and a constant for $\alpha=1$, in agreement with Fig.~\ref{fig:pd_iso}(b). Note also that $\tau_L$ is now of order $t_1^{\alpha}t_2^{1-\alpha}$. This is determined by the power-law region between $t_1$ and $t_2$ in the FPT distribution that still exists for $\alpha>0$, as discussed in Appendix A. 

\subsubsection{Preferences during the diffusion process}
It is interesting to compare the preferences of the particles to be absorbed by a particular wall (the \textit {ultimate preferential direction} at the end of the trajectory)
to their preference for a particular side at earlier times.
The problem can be formulated in different ways.

\begin{figure}[h]
 	\centering
	\includegraphics[width=0.4\textwidth]{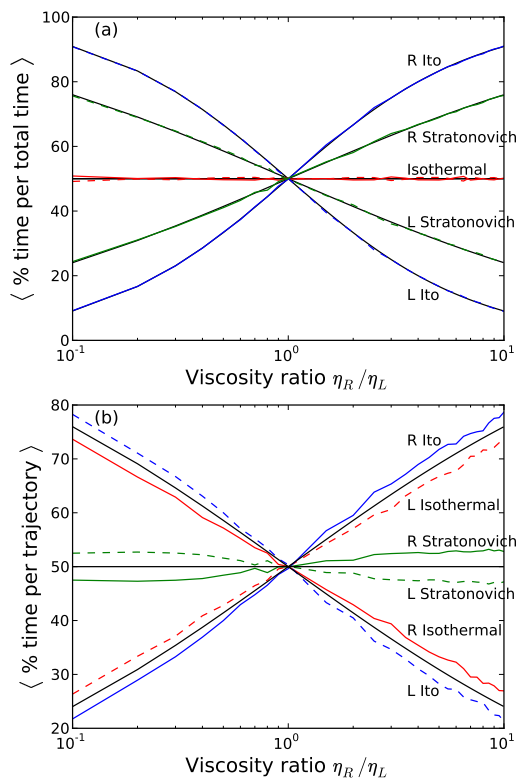}
	\caption{(color online) For the case of absorbing walls with $\eta_L=10$, $d=50$, and varying $\eta_R$, for Ito, Stratonovich, and isothermal calculi, fractions of time spent by trajectories on L and R defined in two different ways, as described in the text: 
	(a) $f_{L,R}$; 
	(b) $\phi_{L,R}$. 
	Black lines are analytic dependences described in the text: Eqs.~(\ref{fL}) and (\ref{fR}) in (a) and Eqs.~(\ref{phiL}) and (\ref{phiR}) in (b). 
	In all cases, particles start at the interface in the middle of the system.}
	\label{fig:frac1}
\end{figure}

First, we can take a sample of particles, calculate the total combined time spent by them on a particular side, and divide by the total combined duration of all trajectories. Obviously, this is the same as the average time that a trajectory spends on a particular side divided by the total average duration of the trajectory. Using again the MC I algorithm with all moves successful, we note that the average number of steps in each segment is the same for left and right segments. Denoting the average total number of steps $n$ and given that the fraction of left segments is $p'^0_{\I -}$, as discussed in the previous section, the average time a trajectory spends on L is
\begin{equation}
t_L=n p'^0_{\I -} t_{\I w}^L,
\end{equation}
similarly the average time on R is
\begin{equation}
t_R=n p'^0_{\I +} t_{\I w}^R,
\end{equation}
and then the fractions of the time on L and R are
\begin{eqnarray}
f_L&=&\frac{t_L}{t_L+t_R}\nonumber\\
&=&\frac{p'^0_{\I -} t_{\I w}^L}{p'^0_{\I -} t_{\I w}^L+p'^0_{\I +} t_{\I w}^R}=\frac{1}{1+(\eta_R/\eta_L)^{1-\alpha}},\label{fL}\\
f_R&=&\frac{t_R}{t_L+t_R}\nonumber\\
&=&\frac{p'^0_{\I +} t_{\I w}^R}{p'^0_{\I -} t_{\I w}^L+p'^0_{\I +} t_{\I w}^R}=\frac{(\eta_R/\eta_L)^{1-\alpha}}{1+(\eta_R/\eta_L)^{1-\alpha}}.\label{fR}
\end{eqnarray}
These functions for $\alpha=0$, 0.5, and 1 are plotted in Fig.~\ref{fig:frac1}(a) together with the results of MC I simulations in these cases, and agreement is excellent. Note that these probabilities are different from the probabilities to reach a particular wall, $\pi_{L,R}$, given by Eqs.~(\ref{piL}) and (\ref{piR}). In particular, while for Ito $\pi_L=\pi_R$, $f_L<f_R$ when $\eta_L<\eta_R$. 
Conversely, for isothermal $f_L=f_R$, but $\pi_L>\pi_R$ for $\eta_L<\eta_R$. In fact, $\pi_L$ for Ito equals $f_R$ for isothermal, and vice versa, or, in general,
\begin{eqnarray}
f_R(\eta_R/\eta_L,\alpha)&=&\pi_L(\eta_R/\eta_L,1-\alpha),\\
f_L(\eta_R/\eta_L,\alpha)&=&\pi_R(\eta_R/\eta_L,1-\alpha).
\end{eqnarray}
At the same time, $f_{L,R}$ coincide with the long-time probabilities for particles to be found on a particular side in the problem with reflecting boundaries [cf. Eq.~(\ref{longtime})].

Second, we can also determine the fraction of the time spent on a particular side for each trajectory in the sample and then \textit{average these fractions}. We denote these averaged fractions $\phi_{L,R}$. This definition gives more weight to short trajectories than the previous one, since a short trajectory contributes little to both the total duration of trajectories and to the duration on a particular side, the quantities that enter the definitions for $f_{L,R}$, but it contributes as much as long trajectories to $\phi_{L,R}$. 
The results for $\phi_{L,R}$ obtained in MC I simulations for $\alpha=0$, 1/2, and 1 are shown in Fig.~\ref{fig:frac1}(b). These results can be fitted using the approximate expressions,
\begin{eqnarray}
\phi_L&\approx&\frac{1}{1+(\eta_R/\eta_L)^{1/2-\alpha}},\label{phiL}\\
\phi_R&\approx&\frac{(\eta_R/\eta_L)^{1/2-\alpha}}{1+(\eta_R/\eta_L)^{1/2-\alpha}}.\label{phiR}
\end{eqnarray}
If the equalities (\ref{phiL}) and (\ref{phiR}) were exact, then, (1) $\phi_L$ for Ito would be equal to $\phi_R$ for isothermal, and (2) $\phi_L$ and $\phi_R$ for Stratonovich ($\alpha=1/2$) would be exactly $1/2$ for all $\eta_R/\eta_L$. \
However, we have checked using several different lattice sizes and extrapolating to the infinite size (results not shown) that these relations are indeed approximate in the continuum limit and both statements (1) and (2) are approximate as well. 

Comparing Eqs.~(\ref{piL}), (\ref{piR}), (\ref{fL}), (\ref{fR}), (\ref{phiL}), and (\ref{phiR}), we note that the average fractions of time spent on the low-viscosity side calculated in both ways are always smaller than the probability to be absorbed by the wall on the low-viscosity side, but the second way of calculating the average gives a result intermediate between the other two. That is, for $\eta_R/\eta_L>1$,
\begin{equation}
f_L<\phi_L<\pi_L.\label{relation}
\end{equation}
The differences between these quantities are particularly striking when the viscosity ratio is large, in which case both ``$<$'' signs in Eq.~(\ref{relation}) turn into ``$\ll$'', so there are ranges of $\alpha$ where one or two of the quantities $f_L$ and $\phi_L$ approach zero while $\pi_L$ approaches unity. These relations are easy to explain by recalling that many particles that are eventually absorbed by the wall on the low-viscosity side linger until time $\gg t_1$ (up to $\sim t_2$) necessarily spending much of that time on the high-viscosity side, which explains why $f_L,\phi_L<\pi_L$. In addition, such long trajectories have a relatively larger weight when calculating $f_L$ than when calculating $\phi_L$, which explains $f_L<\phi_L$. Note that nearly all trajectories reaching the wall on the high-viscosity side spend most of the time on that side.

\section{Discussion}

In this paper, we have developed two dynamical Monte Carlo algorithms for simulating the dynamics of a particle in the presence of sharp viscosity (or, in general, diffusivity) changes applicable to cases where different interpretations of the stochastic term in the corresponding overdamped Langevin equation (or calculi, such as Ito, Stratonovich, and isothermal) are appropriate. 
The validity of these algorithms was demonstrated by the good agreement between the approaches and with Brownian dynamics for the Ito calculus and Langevin dynamics for the isothermal calculus, as well as with theoretical predictions. 
We have argued that MC approaches, such as those developed here, can be particularly advantageous when a sharp interface is present in the system, since solving the overdamped Langevin equation directly is problematic in this case, except for the Ito calculus (corresponding to BD), while the LD approach is only applicable to the isothermal case and generally more computationally intensive.

While our simulations of particles on a 1D interval with the interface in the middle were intended primarily to validate our approaches, the system itself turned out to be rather interesting, with sometimes non-obvious properties that depend significantly on the choice of the calculus characterized by the parameter $\alpha$ that can vary between zero (Ito) and one (isothermal). 
The differences between these extreme cases are summarized in Table~\ref{tab:summary} for a number of features of the system. 
Not only do the features vary greatly between calculi, but there are unexpected symmetries between certain features (indicated by arrows in the table).
Clearly, when modeling a physical system, the choice of calculus in deciding how to treat the interface must be done with great care.

\begin{table*}
\centering
	\includegraphics[width=0.9\textwidth]{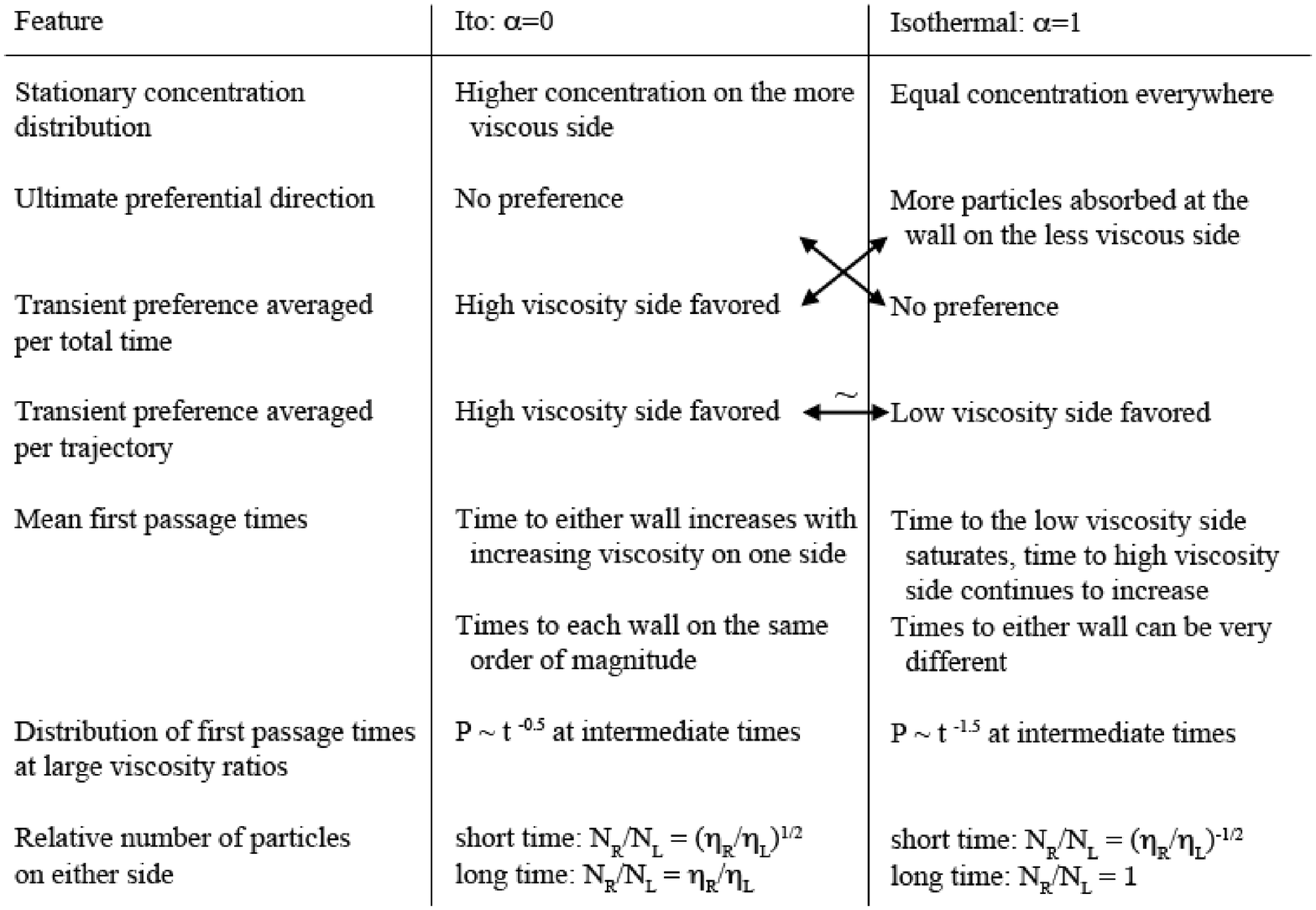}

\caption{
Results for selected features of the system for the two extreme $\alpha$ cases: $\alpha=0$ (Ito calculus) and $\alpha=1$ (isothermal calculus).
The arrows indicate effects of equal magnitude (or approximately so indicated by $\sim$).}
\label{tab:summary}
\end{table*}

The goal for the future is extending the approaches presented here to more complex and physically interesting situations, such as two- and three-dimensional systems and biased diffusion under the action of a force. 
We currently work on applying these algorithms to studying the translocation of a polymer through a hole in a membrane, with fluids of different viscosities on the two sides of the membrane.
Of course, extension to the case of several interfaces is straightforward and a gradual change in viscosity can most likely be modeled as many interfaces one lattice site apart. 
Besides this, even in the simplest case considered here, the algorithms can be further improved: There are several free parameters that we have chosen arbitrarily, but that can be optimized to speed up the convergence to continuum, so even coarse discretization produces accurate results.

It is useful to note that the physical scenario examined in this paper actually belongs to a more general class of problems.
Of course, the discrimination between different calculi
arises for any stochastically driven process that can be written as an overdamped Langevin equation in which the prefactor of the noise term is state-dependent \cite{vankampen1981, lau2007, sokolov2010, coffey2005, vankampen1992, romanczuk2011}.
Hence, while we have focused on diffusion in an inhomogenous medium, similar concepts can be applied to many problems including 
financial modeling \cite{sethi1981, perello2000},
NMR spectroscopy \cite{bartholdi1976},
population growth models \cite{braumann2007}, and climate modeling \cite{majda2009, garcia2011, ewald2003}.
Another interesting example is the work by Smythe \textit{et al.} \cite{smythe1983} examining noise-induced phase transitions where it was found that 
digital simulations produced results in agreement with the Ito formulation while apparently equivalent analog simulations produced results that agreed with the Stratonovich formulation.
Given these diverse scenarios, the simulation approaches introduced herein may be useful outside of the problem for which they were originally intended.

\section{Acknowledgements}

This research was supported by grants from NSERC and the University of Ottawa. Simulations were performed on the SHARCNET computer system.

\section*{Appendix A: Particle distributions and first-passage-time distributions at intermediate times in the high-viscosity-ratio limit}

\setcounter{equation}{0}
\setcounter{figure}{0}
\renewcommand{\theequation}{A.\arabic{equation}}
\renewcommand{\thefigure}{A.\arabic{figure}}

To explain the power-law dependence of the FPT distribution in the intermediate-time regime between $\sim t_1$ [Eq.~(\ref{t1})] and $\sim t_2$ [Eq.~(\ref{t2})], as observed in Fig.~\ref{fig:distLR}, consider the particle distribution function $\rho(x,t)$ in this regime for the problem with absorbing boundaries. Similar to the case of reflecting boundaries discussed in Sec.~\ref{refl}, $\rho(x,t)$ consists of a quasistationary distribution in the low-viscosity region and a ``Gaussian-like'' distribution in the high-viscosity region. (The exact meaning of ``Gaussian-like'' will be revealed later.) The difference is that because of the absorbing boundary conditions [$\rho(x,t)=0$ at the walls], the quasistationary part is no longer a constant, but instead a linear function,
\begin{equation}
\rho(x,t)\approx A(t)(1+2x/d),\ x<0,\label{qst_abs}
\end{equation}
where $A(t)=\rho(-0,t)$. Here and in what follows we assume that the left half of the system is the low-viscosity region ($\eta_L\ll\eta_R$). Equation~(\ref{qst_abs}) provides a relation between the value of $\rho(x,t)$ at $-0$ and its derivative:
\begin{equation}
\left.\frac{\partial\rho(x,t)}{\partial x}\right|_{x=-0} \approx \frac{2}{d}\rho(-0,t).\label{BCleft}
\end{equation}
The slope, $\left.[\partial\rho(x,t)/\partial x]\right|_{x=-d/2}\approx\left.[\partial\rho(x,t)/\partial x]\right|_{x=-0}$, is related to the flux to the left wall and thus to the conditional FPT distribution for the left wall,
\begin{equation}
P_L(t)\approx \frac{D_L}{\pi_L}\left.\frac{\partial\rho(x,t)}{\partial x}\right|_{x=-0}\approx\frac{2D_L}{\pi_L d}A(t).\label{PL}
\end{equation}

From Eq.~(\ref{BCleft}), using the jump condition (\ref{jump}) and the flux continuity condition (\ref{fluxcont}), we obtain a similar relation on the high-viscosity side of the interface:
\begin{equation}
\left.\frac{\partial\rho(x,t)}{\partial x}\right|_{x=+0} \approx \frac{2}{d}\left(\frac{\eta_R}{\eta_L}\right)^{\alpha}\rho(+0,t).\label{RobinBC}
\end{equation}
For the diffusion problem in the high-viscosity region, this serves as a boundary condition (BC). 
BCs of this type, where the derivative is proportional to the value of the function, are known as \textit{Robin boundary conditions} \cite{salsa2008}. 
At times $t\ll t_2$ the right boundary can be ignored, and we can treat this problem as being defined on a semi-infinite interval,
\begin{equation}
\frac{\partial\rho(x,t)}{\partial t}=D_R\frac{\partial^2\rho(x,t)}{\partial x^2},\ x>0,\label{diffR}
\end{equation}
with the boundary condition (\ref{RobinBC}).

The boundary-value problem given by Eqs.~(\ref{diffR}) and (\ref{RobinBC}) can be solved exactly in quadratures \cite{polyanin2002}. We will opt instead for a more intuitive approach, and from now on, we will omit all numerical constants of order unity from our considerations. Since the BC (\ref{RobinBC}) is intermediate between the absorbing BC $\rho(+0,t)=0$ and the reflecting BC $\left.[\partial\rho(x,t)/\partial x]\right|_{x=+0}=0$, we can expect that in some limits the former can be approximated by one of the latter two. For the reflecting BC, with particles starting near $+0$ at time $t=0$, the solution is
\begin{equation}
\rho(x,t)=\frac{B}{\sqrt{t}}\exp\left(-\frac{x^2}{4D_R t}\right),\label{reflsol}
\end{equation}
where $B$ is a constant. At a given time, the maximum value of this function is at $x=0$ and equals $B$; the $x$ derivative at that point is zero. On the other hand, the maximum absolute value of the $x$ derivative is at $x\sim\sqrt{D_R t}$ and is $\sim B/(t\sqrt{D_R})\sim B\sqrt{\eta_R}/t$. The ratio between the maximum absolute value of the derivative and the maximum value of the function is $\sqrt{\eta_R/t}$. If this value is much larger than the ratio of the values of the derivative and the function at $+0$ given by Eq.~(\ref{RobinBC}) [this latter ratio is $\sim (1/d)(\eta_R/\eta_L)^{\alpha}$], then just a small shift in space of the solution of the reflecting problem, Eq.~(\ref{reflsol}), will be sufficient to satisfy the BC (\ref{RobinBC}), so in this situation, which is observed when $\sqrt{\eta_R/t}\gg (1/d)(\eta_R/\eta_L)^{\alpha}$, or for
\begin{equation}
t\ll t_c=\eta_R d^2 \left(\frac{\eta_L}{\eta_R}\right)^{2\alpha}\sim t_1^{2\alpha}t_2^{1-2\alpha},\label{tc}
\end{equation}
the solution of the reflecting problem, Eq.~(\ref{reflsol}), is an adequate approximation of the solution of the original problem with the BC (\ref{RobinBC}). On the other hand, considering the solution of the absorbing problem,
\begin{equation}
\rho(x,t)=\frac{Cx}{t^{3/2}}\exp\left(-\frac{x^2}{4D_R t}\right),\label{abssol}
\end{equation}
where $C$ is a constant, the maximum value of the function is $\sim C\sqrt{D_R}/t\sim C/(t\sqrt{\eta_R})$, the maximum value of the derivative is $C/t^{3/2}$, and the ratio of the latter and the former is $\sqrt{\eta_R/t}$, as in the reflecting case. 
But now, in order for the solution with the BC (\ref{RobinBC}) to be obtainable by a small perturbation of that of the absorbing problem, we need this ratio to be much \textit{smaller} than that specified by Eq.~(\ref{RobinBC}), which happens for $t\gg t_c$, with $t_c$ still given by Eq.~(\ref{tc}). Thus, the solution of the Robin boundary-value problem (\ref{diffR})--(\ref{RobinBC}) crosses over from being ``reflecting-like'' at $t\ll t_c$ to being ``absorbing-like'' at $t\gg t_c$. Note that the ``reflecting-like'' solution corresponds to the particle number on R staying nearly constant, whereas the ``absorbing-like'' solution corresponds to this particle number decreasing rapidly (as $t^{-1/2}$) as the particles are ``absorbed'' by the interface, ``transferred'' to the left side, and ``flow'' to the wall.

\begin{figure}[h!]
 	\centering
	\includegraphics[width=0.4\textwidth]{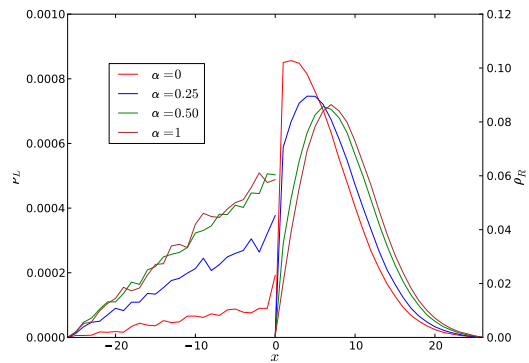}
	\caption{(color online) The particle distribution functions for $\eta_L=2$, $\eta_R=2000$, and $t=10^5$ with absorbing boundary conditions obtained using the variant of the MC I algorithm with the waiting time drawn from distribution (\ref{ndistr}) using $10^6$ realizations. All particles start at the interface in the middle of the system. Different scales are used for the left and right parts of the plot: The left axis corresponds to the left part (with $\eta_L=2$) and the right axis corresponds to the right part (with $\eta_R=2000$). The distributions are normalized so that the area under the curves is unity, regardless of the fraction of particles left.}
	\label{fig:abs_dist}
\end{figure}

However, it should be remembered that our consideration is only valid in the interval $t_1\ll t\ll t_2$. When $\alpha=0$, $t_c\sim t_2$, so the ``reflecting-like'' solution should be observed in the whole range of validity, although some ``traces'' of the ``absorbing-like'' solution can appear as one approaches $t_2$. On the other hand, when $\alpha>1/2$, $t_c<t_1$, so the solution is ``absorbing-like'' in the whole interval. Of course, since for $t<t_1$ the particles do not yet feel the walls, the distribution should be the same as in the reflecting case, that is, ``reflecting-like''; the transition should occur around $t_1$. It is only for $0<\alpha<1/2$ that the crossover should actually be observed in the interval $t_1\ll t\ll t_2$. If we choose $t\sim\sqrt{t_1 t_2}$, then for $\alpha=0$ the particle distribution in the high-viscosity region should resemble Eq.~(\ref{reflsol}); for $\alpha=0.5$ and $\alpha=1$ it should be similar to Eq.~(\ref{abssol}); finally, for $\alpha=0.25$ this time equals $t_c$ and the distribution should be a ``hybrid'' of these two cases. Figure~\ref{fig:abs_dist} showing the distributions for $\eta_L=2$, $\eta_R=2000$, $t=10^5$ indeed matches these expectations. Even though $t=10^5$ is about 2.5 times larger than $\sqrt{t_1 t_2}\approx 40000$ (this larger time is chosen to have broader distributions, so their features are more visible), the crossover is broad enough that this time can still be considered to correspond to the crossover for $\alpha=0.25$. Note that the distributions in the low-viscosity region are linear, as given by Eq.~(\ref{qst_abs}); note also that a different scale had to be used for the low-viscosity region, since the values of the particle density in that region are much lower than in the high-viscosity region. Thus, for any $\alpha$, the fraction of particles in the low-viscosity region is very small in this time range; contrast this with the reflecting case, where, for example, in the isothermal case there are more particles in the low-viscosity region than in the high-viscosity region at all times (except when $t\to\infty$, these numbers become equal). To obtain the plot in Fig.~\ref{fig:abs_dist}, the variant of the MC I algorithm with the waiting time drawn from the distribution (\ref{ndistr}) is used providing significant computational time savings for this large viscosity ratio.

\begin{figure}[h!]
 	\centering
	\includegraphics[width=0.4\textwidth]{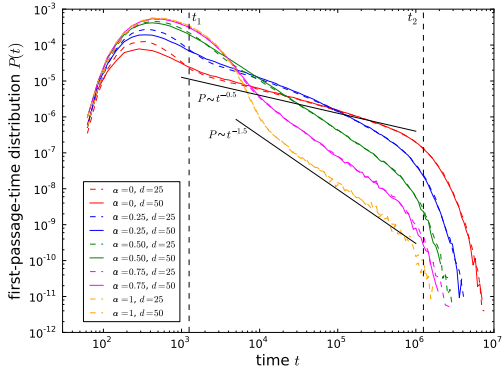}
	\caption{(color online) The FPT distributions for $\eta_L=2$, $\eta_R=2000$, and $d=50$ (dashed lines). All particles start at the interface in the middle of the system. The solid lines are the data for $d=100$, rescaled appropriately (i.e., the time is divided by 4 and the value is multiplied by 4). The data are obtained using the variant of the MC I algorithm with the waiting time drawn from the distribution (\ref{ndistr}) using $10^6$ realizations.}
	\label{fig:dist_tauL}
\end{figure}

We are now in a position to obtain the FPT distribution for the left wall, $P_L(t)$. In the ``reflecting-like'' situation, considering the solution with the reflecting BC, Eq.~(\ref{reflsol}), by itself, the $x$ derivative at $+0$ is, of course, zero. The next iteration is obtained by inserting the value of the function at this point, $B/\sqrt{t}$, into the BC (\ref{RobinBC}), which gives $\sim B(\eta_R/\eta_L)^{\alpha}/(d\sqrt{t})$. Using Eqs.~(\ref{fluxcont}) and (\ref{PL}),
\begin{equation}
P_L(t)\sim \frac{BD_R(\eta_R/\eta_L)^{\alpha}}{d\sqrt{t}}\sim \frac{B\eta_R^{\alpha-1}}{d\eta_L^{\alpha} t^{1/2}}.\label{PLrefl}
\end{equation}
Here we have used the fact that $\pi_L$ is always $\sim 1$. In the ``absorbing-like'' case, the $x$ derivative at $+0$ is $C/t^{3/2}$, and in the same way we obtain
\begin{equation}
P_L(t)\sim \frac{C}{\eta_R t^{3/2}}.\label{PLabs}
\end{equation}
Thus, the $P(t)$ dependence should always be a power law; the exponent can be $-1/2$ in the ``reflecting-like'' situation or $-3/2$ in the ``absorbing-like'' situation. In Fig.~\ref{fig:dist_tauL}, we plot $P_L(t)$ for $\eta_L=2$ and $\eta_R=2000$. We see that indeed, for $0.5\le\alpha\le 1$ (which corresponds to the ``absorbing-like'' situation at all times), the dependence is $\sim t^{-3/2}$. For $\alpha=0$ (the ``reflecting-like'' case), the $t^{-1/2}$ fit is adequate, although the slope does get steeper close to $t_2$, as the crossover is starting to be ``felt.'' For a smaller viscosity ratio, as in Fig.~\ref{fig:distLR}, the crossover is felt over the whole intermediate range of times, which is probably what causes the apparent $t^{-0.8}$ dependence in that case. Finally, for $\alpha=1/4$ the curve cannot be fitted by a simple power-law function; apparently, the crossover is so broad that the already large viscosity ratio needs to be increased by at least another order of magnitude before the distinct $t^{-1/2}$ and $t^{-3/2}$ regions can be observed.

To obtain the prefactors (or the coefficients $B$ and $C$) in $P_L(t)$, we assume that the \textit{number} of particles on R, $N_R$, at the beginning of the intermediate time interval ($t\sim t_1$) is still the same as at $t\to 0$, which, in its turn, is the same as in the reflecting case, since at $t\to 0$ the particles have not reached the boundaries yet. Note that this does not mean that the \textit{ratio} $N_R/N_L$ at $t\sim t_1$ is the same as at $t\to 0$, because particles on L do start to be absorbed around $t_1$, so $N_L$ does change by $\sim t_1$. This means that while Eq.~(\ref{shorttime}) can be used to get $N_R$ at $t\sim t_1$, $N_L$ needs to be replaced with $N_0-N_R$, where $N_0$ is the total \textit{initial} number of particles on both sides. Two cases need to be considered. For $\alpha<1/2$, Eq.~(\ref{shorttime}) (modified as above) gives $N_R\gg N_0-N_R$, so
\begin{equation}
N_R/N_0\sim 1.\label{NRlesshalf}
\end{equation}
On the other hand, for $\alpha\ge 1/2$,
\begin{equation}
N_R/N_0\sim (\eta_R/\eta_L)^{1/2-\alpha}\ll 1.\label{NRgtrhalf}
\end{equation}
The first case is ``reflecting-like'' at $t_1$ and until $t_c$; normalizing the particle distribution function so the area under the whole curve (L and R) is unity initially [and therefore, given Eq.~(\ref{NRlesshalf}), the area under R is still approximately unity at $\sim t_1$], we obtain $(B/\sqrt{t_1})\sqrt{D_R t_1}\sim 1$, thus $B\sim 1/\sqrt{D_R}\sim\sqrt{\eta_R}$, and using Eq.~(\ref{PLrefl}),
\begin{equation}
P_L(t;\alpha<1/2)\sim\frac{\eta_R^{\alpha-1/2}}{d\eta_L^{\alpha}t^{1/2}}\sim\frac{t_2^{\alpha-1/2}}{t_1^{\alpha}t^{1/2}},\ t\lesssim t_c.\label{PL11}
\end{equation}
After the crossover, $P_L(t)\propto t^{-3/2}$; by matching to Eq.~(\ref{PL11}) at the crossover point $t_c$,
\begin{equation}
P_L(t;\alpha<1/2)\sim\frac{t_1^{\alpha}t_2^{1/2-\alpha}}{t^{3/2}},\ t\gtrsim t_c.\label{PL12}
\end{equation}
The case $\alpha\ge 1/2$ is ``absorbing-like'' for all intermediate $t$; the area under the curve given by Eq.~(\ref{abssol}) at $t=t_1$ is $\sim CD_R/t_1^{1/2}\sim C/(\eta_R t_1^{1/2})$, so, according to Eq.~(\ref{NRgtrhalf}), the normalization condition is $C/(\eta_R t_1^{1/2})\sim (\eta_R/\eta_L)^{1/2-\alpha}$, thus $C\sim \eta_R^{3/2-\alpha}t_1^{1/2}/\eta_L^{1/2-\alpha}$, and, using Eq.~(\ref{PLabs}),
\begin{equation}
P_L(t;\alpha>1/2)\sim\frac{t_1^{\alpha}t_2^{1/2-\alpha}}{t^{3/2}},\label{PL2}
\end{equation}
which is the same expression as Eq.~(\ref{PL12}).

Note that Eqs.~(\ref{PL11}) and (\ref{PL2}) are not necessarily valid, even approximately, at $t\approx t_1$, mostly because around $t_1$, and until time $t^*=\beta t_1$, where $\beta$ is a sufficiently large factor, the particle distribution on L may still be very different from that given by Eq.~(\ref{qst_abs}) . [Strictly speaking, this means that using $t\sim t_1$ in considerations leading up to Eqs.~(\ref{PL11})--(\ref{PL2}) is not entirely correct, but since for all practical purposes $\beta$ can be chosen as a large, but finite and constant numerical factor, this does not affect the scalings, apart from at most logarithmic corrections.] For $\alpha>1/2$, the number of particles on $R$ relative to the total initial number of particles in the system $N_0$ is always $\ll 1$. On the other hand, using Eq.~(\ref{PL}), the number of particles on $L$ (again, relative to $N_0$) at $t=t^*$ (at which the above considerations are already valid but still as close to $t_1$ as possible) is
\begin{equation}
N_L(t^*)\sim A(t^*)d\sim \frac{P_L(t^*)d^2}{D_L}\lesssim \left(\frac{t_1}{t_2}\right)^{\alpha-1/2}\ll 1.
\end{equation}
Thus by time $t^*$ most particles have already disappeared from the system. This absorption of particles must have happened around time $\sim t_1$ over the time interval $\sim t_1$. Thus, $P_L(t_1)\sim 1/t_1$. However, Eq.~(\ref{PL2}) predicts $P_L(t_1)\sim (1/t_1)(t_1/t_2)^{\alpha-1/2}\ll 1/t_1$. So, on top of the power law predicted by Eq.~(\ref{PL2}) there should be an additional peak (or ``bulge'') at $\sim t_1$ of height $\sim 1/t_1$:
\begin{equation}
P_L(t_1;\alpha>1/2)\sim 1/t_1.
\end{equation}
This ``bulge'', in fact, contains nearly all particles, although the rest of the distribution is still important when calculating the MFPT, as we will see. On the other hand, for $\alpha<1/2$ only a small fraction of all particles, $\sim (\eta_R/\eta_L)^{\alpha-1/2}$, are on L initially. Even if all of them are absorbed over time $\sim t_1$, the resulting FPT density is $\sim (\eta_R/\eta_L)^{\alpha-1/2}/t_1\sim t_2^{\alpha-1/2}/t_1^{\alpha+1/2}$, which is the same as what Eq.~(\ref{PL11}) at $t=t_1$ gives, so there is no additional ``bulge". In fact, in Fig.~\ref{fig:dist_tauL} there appear to be ``bulges'' for all $\alpha$. However, it should be kept in mind that our discretization is not entirely adequate for this problem with a very high viscosity ratio: at $t=t_1=1250$, the width of the distribution on R is less than one, that is, less than one mesh step. In the same figure, we also show the results obtained for $d=100$, with the resulting distributions appropriately rescaled. The ``bulges'' on the corresponding curves with $\alpha<1/2$ are smaller, so it is reasonable to assume that they largely disappear in the continuum limit. On the other hand, the ``bulges" on curves with $\alpha>1/2$ are of the same size for $d=50$ and $d=100$ and therefore are ``real'', rather than discretization artifacts.

Using our results for FPT, we can now analyze the expression for the MFPT we have obtained [Eq.~(\ref{tauLgenlim})], by decomposing the MFPT into parts associated with different regions of the FPT distribution. For $\alpha>1/2$, the weight of the bulge is nearly one, while the weight of the power-law region is
\begin{equation}
\int_{t_1}^{t_2}P_L(t)dt\sim\int_{t_1}^{t_2} \frac{t_1^{\alpha}t_2^{1/2-\alpha}}{t^{3/2}}dt\sim\left(\frac{t_1}{t_2}\right)^{\alpha-1/2}\ll 1.
\end{equation}
On the other hand, the average time over the ``bulge" is $\sim t_1$, but the average time over the power-law region is
\begin{equation}
\frac{\int_{t_1}^{t_2}tP_L(t)dt}{\int_{t_1}^{t_2}P_L(t)dt}\sim\frac{\int_{t_1}^{t_2}t^{-1/2}dt}{\int_{t_1}^{t_2}t^{-3/2}dt}\sim t_1^{1/2} t_2^{1/2},
\end{equation}
which is much larger. The total MFPT is
\begin{equation}
\tau_L\sim 1\times t_1+\left(\frac{t_1}{t_2}\right)^{\alpha-1/2}\times (t_1^{1/2}t_2^{1/2})\sim t_1+t_1^{\alpha}t_2^{1-\alpha},
\end{equation}
where the first term comes from the ``bulge'' and the second one comes from the power-law region at intermediate times. Note that the second term is always larger than the first one (except for $\alpha=1$, when they are equal), and therefore it is the second term that determines the character of the MFPT dependence, which indeed coincides with Eq.~(\ref{tauLgenlim}). For $\alpha<1/2$, there are two power-law regions. The weight of the first, ``reflecting-like'' region is
\begin{equation}
\int_{t_1}^{t_c}\frac{t_2^{\alpha-1/2}}{t_1^{\alpha}t^{1/2}}dt\sim \frac{t_2^{\alpha-1/2}t_c^{1/2}}{t_1^{\alpha}}\sim 1.
\end{equation}
The weight of the second, ``absorbing-like'' region is
\begin{equation}
\int_{t_c}^{t_2}\frac{t_1^{\alpha}t_2^{1/2-\alpha}}{t^{3/2}}dt\sim \frac{t_1^{\alpha}t_2^{1/2-\alpha}}{t_c^{1/2}}\sim 1.
\end{equation}
So both regions carry equal weight. In fact, the largest contribution comes from the region around $t_c$ --- the same result is obtained if the lower limit of the first integral and the upper limit of the second integral are replaced by arbitrary values, as long as the first one is much smaller than $t_c$ and the second one is much larger than $t_c$. To put it differently, of all intervals $[t_a,t_b]$ with equal $t_b/t_a$, the one containing $t_c$ will contribute the most. However, other regions cannot be ignored when calculating the MFPT. The average time over the ``reflecting-like'' region is
\begin{equation}
\frac{\int_{t_1}^{t_c} t^{1/2}dt}{\int_{t_1}^{t_c} t^{-1/2}dt}\sim t_c\sim t_1^{2\alpha}t_2^{1-2\alpha},
\end{equation}
while that over the ``absorbing-like'' region is
\begin{equation}
\frac{\int_{t_c}^{t_2} t^{-1/2}dt}{\int_{t_c}^{t_2} t^{-3/2}dt}\sim t_c^{1/2}t_2^{1/2}\sim t_1^{\alpha}t_2^{1-\alpha}.
\end{equation}
The resulting MFPT is
\begin{equation}
\tau_L\sim 1\times t_1^{2\alpha}t_2^{1-2\alpha}+1\times t_1^{\alpha}t_2^{1-\alpha}.
\end{equation}
The contribution of the second term dominates (except for $\alpha=0$, when these contributions are equal, even though the second region ``shrinks into a point"). 
Again, this second term coincides with Eq.~(\ref{tauLgenlim}). Note that for both $\alpha<1/2$ and $\alpha>1/2$, contributions of the parts of the distribution with $t>t_1$ are significant. Particles spending such a long time before being absorbed necessarily spend most of that time in the high-viscosity region. Note in this regard that the term in Eq.~(\ref{tauLgen}) associated with the time spent on R dominates for $0\le\alpha<1$ when $\eta_R/\eta_L\to\infty$, and even at $\alpha=1$ the terms corresponding to the time spent on L and on R contribute equally (by order of magnitude).

It is interesting to compare the particle distributions in Fig.~\ref{fig:abs_dist} with those obtained for the same $\eta_L$, $\eta_R$, $d$, and $t$ for reflecting boundaries. The crucial difference is that, while in the absorbing case the flow of particles is from the high-viscosity to the low-viscosity region, the opposite is true in the reflecting case, which follows from the fact that for $\eta_R/\eta_L>1$ and for all $\alpha$, the ratio $N_R/N_L$ grows with time [see Fig.~\ref{fig:dist_all}(c)], yet $N_L + N_R$ stays constant. As mentioned, this growth is proportional to $\sqrt{t}$; given that it occurs between $t_1$ and $t_2$ and using the short-time [Eq.~(\ref{shorttime})] and long-time [Eq.~(\ref{longtime})] values of this ratio, we get
\begin{equation}
\frac{N_R}{N_L}\sim \frac{t_2^{1/2-\alpha}}{t_1^{1-\alpha}}t^{1/2}.\label{NRNL}
\end{equation}
For $\alpha<1/2$ this ratio is much larger than unity at all times, and the fraction of particles on R is
\begin{equation}
\frac{N_R}{N_L+N_R}\approx 1-\frac{N_L}{N_R}\approx 1-\frac{t_1^{1-\alpha}}{t_2^{1/2-\alpha}}t^{-1/2}.
\end{equation}
The derivative of this is equal to the flux of particles into the right region, which in its turn is proportional to the space derivative of the particle distribution function at $x=+0$. Thus,
\begin{equation}
\frac{1}{\eta_R}\left.\frac{\partial\rho(x,t)}{\partial x}\right|_{x=+0}\sim -\frac{t_1^{1-\alpha}}{t_2^{1/2-\alpha}}t^{-3/2},
\end{equation}
or
\begin{equation}
\left.\frac{\partial\rho(x,t)}{\partial x}\right|_{x=+0}\sim -\frac{t_1^{1-\alpha}}{t_2^{1/2-\alpha}}\eta_R t^{-3/2}.\label{reflder}
\end{equation}
Assuming that the distribution on R can be approximated by one half of a Gaussian curve with area $\sim 1$ [Eq.~(\ref{reflsol}) with $B\sim 1/\sqrt{D_R}\sim\sqrt{\eta_R}$], the maximum absolute value of the $x$ derivative is $\sim\eta_R/t$. This is much larger than the absolute value of the derivative given by Eq.~(\ref{reflder}) for any $t\gg t_1$. Thus, arguing as in the absorbing case, the Gaussian with the peak at zero [Eq.~(\ref{reflsol})] is indeed a good approximation in this case.

Consider now $\alpha>1/2$. In this case, two situations are possible: for
\begin{equation}
t\gg t_{cr}\sim t_1^{2-2\alpha}t_2^{2\alpha-1},
\end{equation}
$N_R/N_L$ is still large, and the considerations above for $\alpha<1/2$ mostly apply. In particular, Eq.~(\ref{reflder}) and the expression for the maximum absolute value of the $x$ derivative of the trial Gaussian are still valid, and the absolute value of the former is much smaller than the latter for any $t\gg t_{cr}$, so the Gaussian with a peak at zero still approximates the distribution well. On the other hand, for $t\ll t_{cr}$, $N_R/N_L\ll 1$. This means that $N_L$ stays nearly constant and thus $\rho(x,t)$ on L, besides being nearly constant in space, is also nearly constant in time and
\begin{equation}
\rho(x,t)\approx\frac{2}{d}\sim \frac{1}{d},\ x<0,\ t\ll t_{cr}.
\end{equation}
Then, according to the jump condition (\ref{jump}),
\begin{equation}
\rho(x=+0,t)\sim \frac{1}{d}\sqrt{\frac{\eta_R}{\eta_L}}
\end{equation}
and is likewise time-independent. The solution of the diffusion equation satisfying the condition $\rho(+0,t)={\rm const}$ for $t>0$ is the complementary error function erfc \cite{philbert1991}, and we get
\begin{equation}
\rho(x,t)\sim \frac{1}{d}\sqrt{\frac{\eta_R}{\eta_L}}{\rm erfc}\left(\frac{x}{\sqrt{4D_R t}}\right),\ x>0.\label{erfc}
\end{equation}
Note that $N_R\sim\sqrt{t}$, as expected, and also that the inflection point is now at $x=+0$ (indeed, $\partial\rho/\partial t=0$ at $x=+0$ implies $\partial^2\rho/\partial x^2=0$).

\begin{figure}[h!]
 	\centering
	\includegraphics[width=0.4\textwidth]{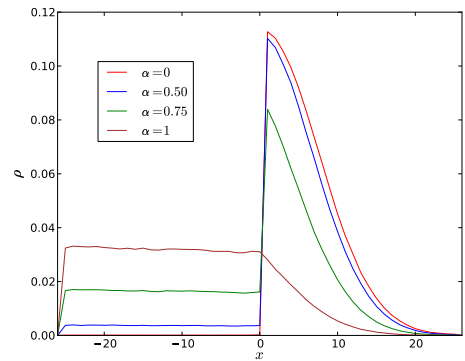}
	\caption{(color online) The particle distribution functions for $\eta_L=2$, $\eta_R=2000$, $t=10^5$ with reflecting boundary conditions obtained using the variant of the MC I algorithm with the waiting time given by Eq.~(\ref{allsucc}) using $10^6$ realizations. All particles start at the interface in the middle of the system.}
	\label{fig:refl_dist}
\end{figure}

Similar to the absorbing case, we can consider the particle distributions at $t\sim\sqrt{t_1 t_2}$. We expect the Gaussian distribution for $\alpha$ between 0 and 0.5, the complementary error function with the inflection point at 0 for $\alpha=1$ and a ``hybrid'' situation for $\alpha=0.75$. In Fig.~\ref{fig:refl_dist}, we show the results obtained using MC I simulations with $t_{\I w}$ given by Eq.~(\ref{allsucc}). These results confirm our expectations. The $\alpha=0.75$ curve resembles the $\alpha=1$ curve, but in the former case the inflection point is clearly shifted to the right of the interface.

Comparing the results in Figs.~\ref{fig:abs_dist} and \ref{fig:refl_dist}, it is interesting to note a kind of ``mirror symmetry'' between them. For example, in the absorbing case the distributions with $\alpha$ between 0.5 and 1 are very similar in the high-viscosity region, and in the reflecting case the distributions with $\alpha$ between 0 and 0.5 have this property. In the absorbing case, $\alpha=0.25$ is ``hybrid,'' while in the reflecting case, $\alpha=0.75$ is ``hybrid.'' Finally, in the absorbing case the distribution for $\alpha=1$ is given by Eq.~(\ref{abssol}), which is the derivative of the Gaussian distribution, and in the reflecting case the distribution for $\alpha=1$ is given by the complementary error function [Eq.~(\ref{erfc})], which is the integral of the Gaussian distribution.

\section*{Appendix B: Proportion of the number of steps on left and right for a random walk absorbed at a particular wall}

\setcounter{equation}{0}
\renewcommand{\theequation}{B.\arabic{equation}}

Using the MC I algorithm with all move attempts successful (Sec.~\ref{MCIallsucc}) reduces the problem of particle diffusion in a system with a diffusivity interface to a lattice random walk that is unbiased everywhere, except perhaps at the interface site. Depending on what side of the interface the particle is on, a single step of this random walk takes a different amount of time, but this can be taken care of separately, by working in terms of the numbers of steps on each side and then multiplying by the respective time steps. 
In this way, our problem is reduced to that of an ordinary lattice random walk with a constant diffusion rate that starts at the site in the middle between two absorbing walls and can be biased at that site, but nowhere else. We first consider the case with no bias even at the middle site (corresponding to the Ito calculus) and then generalize.

First, consider the diffusion on a lattice with mesh step $a=1$ of an ensemble of particles released half-way between walls at $x=\pm d/2$, with $d/2$ integer and the initial position of each particle given by $x_0 = 0$. By symmetry, the average mean first passage times (MFPT) to either wall (the conditional MFPTs) are equal and thus equal to the MFPT to either wall (the unconditional MFPT).
If the time is measured in terms of the number of steps, then the MFPT is the average number of steps, which is \cite{chubynsky2012}
\begin{equation}
n=\frac{d^2}{4}.\label{n}
\end{equation}

To proceed, we can decompose the full trajectory of a particle into segments separated by its visits to the middle site. Each such segment is entirely either in the left half or in the right half and for brevity, we refer to these groups of segments as left and right segments. Consider the ensemble of trajectories ending on the right wall. Taking one such trajectory and replacing all of the right segments (except for the last one, of course) with its mirror-image left segment and vice-versa produces another valid trajectory belonging to the ensemble. Therefore, up to the last segment, there should be an equal number of left and right segments on average and they should have the same average duration. Thus, we can divide the trajectory into two parts. The first part contains all but the last segment and on average, the time spent in both halves of the interval is the same, but the second part, consisting of only the last segment, is entirely in the right half.

The average time corresponding to the second part of the trajectory can be obtained by recasting this problem as an equivalent one. Consider the interval consisting of just the right half of the original interval. If both boundaries of this interval are treated as absorbing, the particle is released right next to the boundary corresponding to the middle site and the conditional MFPT, given that the particle is absorbed by the other boundary, is calculated, then all trajectories which return to the middle site will be discarded. While the probability of being absorbed by a wall vanishes in the limit when the distance from the point where the particle is released to the opposite wall approaches zero (on a lattice this probability is $\sim a/d$), the MFPT to the far wall is well-defined and is, for the continuum problem \cite{redner},
\begin{equation}
\tau_2=\frac{l^2}{6D}=\frac{d^2}{24D},
\end{equation}
where $l=d/2$ is the distance between the walls (the length of the half-interval under consideration). Using Eq.~(\ref{D_LMC}) with $p_s=0$, $a=1$ and $\tau=1$, we convert this result into that for the average number of steps in the corresponding lattice problem,
\begin{equation}
n_2=\frac{d^2}{12}.\label{n2}
\end{equation}
This is the average duration of the second part of the trajectory.
The average duration of the first part is
\begin{equation}
n_1=n-n_2=\frac{d^2}{6}.\label{n1}
\end{equation}
Given that on average, the first part contains an equal number of steps on L and on R, but the second part is entirely on the same side as the wall of interest, the average number of steps on the same side as the wall is
\begin{equation}
n_{\rm same}=\frac{1}{2}n_1+n_2=\frac{d^2}{6},
\end{equation}
and the average number of steps on the opposite side is
\begin{equation}
n_{\rm opp}=\frac{1}{2}n_1=\frac{d^2}{12}.
\end{equation}

Consider now the case when there is bias at the interface, so a particle at the middle site chooses to move left and thus start a left segment with probability $p'^0_{\I -}$ or to move right and thus start a right segment with probability $p'^0_{\I +}=1-p'^0_{\I -}$. These probabilities for MC I are given by Eqs.~(\ref{pminussucc}) and (\ref{pplussucc}). Each of these segments is unbiased and thus, statistically speaking, unmodified compared to the case of no bias. In particular, the distribution of the number of steps in each segment is the same and the probability to reach the wall is the same as well. This means that Eqs.~(\ref{n}), (\ref{n2}), and (\ref{n1}) are still valid. The only thing that changes is the fractions of left and right segments in the first part of the trajectory, which are now $p'^0_{\I -}$ and $p'^0_{\I +}$, respectively. Then, for trajectories ending on the right wall, the average number of steps on R is
\begin{equation}
n_{RR}=p'^0_{\I +} n_1+n_2=\frac{d^2}{12}(1+2p'^0_{\I +}),
\end{equation}
and on L,
\begin{equation}
n_{RL}=p'^0_{\I -} n_1=\frac{d^2}{6}p'^0_{\I -}.
\end{equation}
Similarly, for trajectories ending on the left wall, the average number of steps on L is
\begin{equation}
n_{LL}=p'^0_{\I -} n_1+n_2=\frac{d^2}{12}(1+2p'^0_{\I -}),
\end{equation}
while on R,
\begin{equation}
n_{LR}=p'^0_{\I +} n_1=\frac{d^2}{6}p'^0_{\I +}.
\end{equation}

\end{document}